\documentclass{jaa}
\usepackage{amssymb}
\usepackage{geometry}
\usepackage{multirow}
\usepackage{natbib}
\usepackage{array}
\usepackage{float}
\usepackage{url}
\usepackage{threeparttable}
\usepackage{xtab}
\usepackage{booktabs}
\usepackage{multirow}
\usepackage{longtable}
\usepackage[utf8]{inputenc}
\usepackage{makecell}
\usepackage{microtype}
\usepackage{xcolor}
\usepackage{natbib}
\usepackage{graphicx}
\usepackage{hyperref}
\usepackage{amsmath} 
\usepackage{txfonts} 
\usepackage{siunitx} 
\usepackage{txfonts} 
 \usepackage{ulem}

\begin{document}

\title{NGC 2392 and NGC 4361: Spectroscopic Diagnostics of Planetary Nebula Evolution}


\author{Atul Kumar Singh\textsuperscript{1},  Saurabh Sharma\textsuperscript{2}, Rahul Kumar Anand\textsuperscript{1}, Arpan Ghosh\textsuperscript{3}, Tarak Chand\textsuperscript{2,4} and Shantanu Rastogi\textsuperscript{1,*}}
\affilOne{\textsuperscript{1} Department of Physics, Deen Dayal Upadhyaya Gorakhpur University, Civil Lines, Gorakhpur University -273009\\}
\affilTwo{\textsuperscript{2} Aryabhatta Research Institute of Observational Sciences (ARIES), Manora Peak, Nainital, Uttarakhand -263001\\}
\affilThree{\textsuperscript{3}  Universidad Nacional Aut\'onoma de M\'exico, Instituto de Radioastronom\'ia y Astrof\'isica, Antigua Carretera a P\'atzcuaro 8701, 
Ex-Hda. San Jos\'e de la Huerta, 58089 Morelia, Michoac\'an, M\'exico\\}
\affilFour{\textsuperscript{4}M.J.P. Rohilkhand University Bareilly-243006, India.\\}


\twocolumn[{

\maketitle

\corres{shantanu\_r@hotmail.com}


\begin{abstract}
The study presents a detailed spectroscopic analysis of the planetary nebulae (PNe) NGC~2392 and NGC~4361 using optical spectra obtained from the 2-m Himalayan Chandra Telescope (HCT) and mid-infrared spectra from archival \textit{Spitzer} IRS data. The physical conditions, such as electron temperature ($T_e$) and density ($n_e$), were derived using diagnostic emission lines through the PyNeb software. Elemental abundances of key species, including He, O, N, Ne, S, Cl, and Ar, were determined for both nebulae, offering insights into their nucleosynthesis history and evolutionary status. High-resolution HST and Pan-STARRS imaging further elucidate the morphological structures of the nebulae. NGC~2392 exhibits a well-defined double-shell structure and moderate excitation characteristics, while NGC~4361 displays a diffuse elliptical morphology with high-excitation conditions and a notably low nitrogen content. The observed line spectra and derived abundances point toward distinct progenitor histories for the two PNe, with NGC~2392 originating from a younger, intermediate-mass progenitor, while NGC~4361 traces an older, metal-poor Population II star. This comparative study enhances our understanding of the evolution and chemical enrichment processes of low- to intermediate-mass stars.
\end{abstract}


}]


\doinum{12.3456/s78910-011-012-3}
\artcitid{\#\#\#\#}
\volnum{000}
\year{0000}
\pgrange{1--}
\setcounter{page}{1}
\lp{1}

\section{Intoduction} 

Planetary nebulae (PNe) represent a crucial stage in the life cycle of low- and intermediate-mass stars ($\sim$0.8-8 M$_{\odot}$), occurring just after the asymptotic giant branch (AGB) phase and before the transition to a white dwarf. The study of PNe is fundamental for several reasons. PNe provide invaluable insights into stellar evolution by revealing the processes that govern the transformation from the AGB phase to the white dwarf stage. They play a vital role in galactic chemical evolution by ejecting the outer envelope, contributing to the enrichment of the interstellar medium with heavier elements, such as carbon, nitrogen, and oxygen. The investigation of PNe also enhances our understanding of nucleosynthesis in AGB stars, as these objects offer direct evidence of the products of nuclear reactions and element formation. Thus, PNe serve as essential laboratories for studying stellar processes and the chemical enrichment of galaxies  \citep{ Kwitter_2022, Ventura_2017}.

The central stars of PNe are extremely hot, emitting intense ultraviolet (UV) radiation that ionizes the surrounding gas \citep{Kwok2000}. This ionized gas emits several collisionally excited, bright spectral lines, prominently [O~\textsc{iii}] at 5007\,\AA{} and [N~\textsc{ii}] at 6584\,\AA{} \citep{Osterbrock2006}. These emission lines are crucial for the cooling of PNe; they allow the nebular gas to lose energy through the radiation of photons. Excited-state electrons spontaneously emit specific spectral lines. Thus, removing kinetic energy from the gas, and cooling the nebula to maintain thermal equilibrium \citep{Ferland2003}.

Analyzing the intensities and ratios of these emission lines enables us to probe the physics and chemical compositions of the nebula in detail \citep{Aller1984}. Spectral analysis also helps identify various ions in the nebula and determine the evolution history of the progenitor star \citep{Danehkar2022, GarciaRojas2020}. For instance, the ratio of [O~\textsc{iii}] $\lambda$5007 to H$\beta$ provides information about the oxygen abundance relative to hydrogen \citep{DelgadoInglada2014}. Specific line ratios are also sensitive to electron temperatures and densities; for example, [O III] 4363, 5007; [NII] 5755, (6548+6584); [SIII] 6312, 9068; [NIII] 3869, 3967 and [Cl III] 5517, 5537 ratios are used to derive electron temperatures, while the [S II] 6716, 6731; [OII] 3726, 3739 and [Ar] 4711, 4749 ratio helps determine electron densities \citep{Zhang2005}. These diagnostics allow for a detailed understanding of the physical conditions within the nebula.

Two PNe, NGC 2392 and NGC 4361, which have contrasting properties, are selected for the present study. Both nebulae exhibit rich emission-line spectra and distinctive morphological features, allowing for a comprehensive analysis of nebular physics, chemical abundances, and ionization structures.

NGC 2392, commonly known as the "Eskimo Nebula" is notable for its complex double-shell structure composed of a bright inner shell with the size of $18^{\prime\prime}$ x $15^{\prime\prime}$ and a faint round outer shell with a radius of about $23^{\prime\prime}$. It displays prominent emission lines from various ions, including [O\,\textsc{iii}], [N\,\textsc{ii}], [S\,\textsc{ii}], and He\,\textsc{ii}, enabling detailed analyses of its chemical abundances, electron temperatures, and densities \citep{FangLiu2011, GarciaRojas2012}. The presence of both low- and high-ionization species makes NGC 2392 an ideal laboratory for studying the ionization structure and thermal properties of PNe. Extensive observational data, including high-resolution integral field spectroscopy, can facilitate the testing of theoretical models \citep{Danehkar2018} and enhance our understanding of the processes affecting PN formation and evolution.

On the other hand, NGC 4361 is distinguished by its unusually high ionization levels and complex morphology. It is one of the rare PNe where the emission line of He\,\textsc{ii} at 4686\,\AA{} is stronger than H$\beta$\citep{Torres1990}. The presence of ultra-high-ionization species such as [Ne\,\textsc{v}]\,3426\,\AA{} and a high electron temperature of approximately 18,000\, K \citep{Torres1990} provides a unique opportunity to study the physical conditions in highly ionized PNe. Additionally, its elliptical shape with hook-shaped extensions and evidence of bipolar or quadrupolar outflows make it an interesting object to explore the mechanisms driving nebular shaping and dynamics \citep{Miranda1999, MuthuAnandarao2001}.

Selecting NGC 2392 and NGC 4361 enables us to perform a comparative analysis of PNe that exhibit different ionization levels and morphological features. The diversity among (PNe), particularly in their morphologies, ionization structures, and chemical compositions, is closely linked to the evolution of their central stars. Understanding these relationships is essential for tracing the late stages of stellar evolution and for interpreting the physical and chemical conditions observed in the surrounding nebulae \citep{Balick2002}.

\begin{table}[h!]
\centering
\small
\begin{threeparttable}  
  \caption{Log of HCT spectroscopic observations, date: 2023 April 12}
  \label{tab:observation_log}
  \begin{tabular}{@{}lllcc@{}}
  \toprule
    Grating & Type   & Name      & Exposure                             & Airmass \\
    \midrule
    grism~7 & object & NGC\,2392 & $2\times600$                         & 1.4     \\
    grism~8 & object & NGC\,2392 & $2\times600$                         & 1.5     \\
    grism~7 & object & NGC\,4361 & $2\times600$                         & 2.2     \\
    grism~7 & stds   & Feige\,34 & $1\times720$                         & 1.6     \\
    grism~8 & stds   & Feige\,34 & $1\times720$                         & 1.7     \\
    grism~7 & flat   & Halogen   & $5\times3$                           & ---     \\   
    grism~8 & flat   & Halogen   & $5\times3$                           & ---     \\
    grism~7 & arc    & FeAr      & $3\times12$                          & ---     \\
    grism~8 & arc    & FeNe      & $3\times8$                           & ---     \\
    ---     & bias   & ---       & $12\times0$                          & ---     \\ 
    \bottomrule
  \end{tabular}
  Note: Exposure is given as number of frames\,$\times$\, exposure time in seconds.
  \end{threeparttable}
\end{table}

\section{Observational Data}
\subsection{2m HCT Optical Spectra}
The optical spectra of the target sources (NGC 2392 and NGC 4361) were obtained using a Himalayan Faint Object Spectrograph Camera (HFOSC)\footnote[1]{https://www.iiap.res.in/centers/iao/facilities/hct/hfosc-specs/} attached to a 2-m Himalayan Chandra Telescope (HCT) at the Indian Astronomical Observatory (IAO), Hanle, during 2023. The log of observations is provided in Table \ref{tab:observation_log}. 

HFOSC is an optical imager cum spectrograph used for low- and medium-resolution grism spectroscopy. The grism and the camera combinations used for the observations provide a spectral resolution of $R\sim$ 1330 for grism 7 and $R\sim$ 2190 for grism 8. We have obtained spectra using grisms 7 and 8 that cover the wavelengths from 3800 to 6840 (\text{\AA}) and 5800 to 8350 (\text{\AA}), respectively. The slit was placed through the central stars of the PNe during the observations. Observations of Fe-Ar and Fe-Ne hollow cathode lamps were taken immediately before and after the stellar exposures for wavelength calibrations. Optical standard star Feige 34 (V = 11.18, spectral type: DO) was also observed for the flux calibration.

Data reduction was performed following standard procedures using the IRAF\footnote[2]{IRAF is distributed by the National Optical Astronomical Observatories, which is operated by the Association for Universities for Research in Astronomy, Inc., under contract to the National Science Foundation} software spectroscopic reduction package. One-dimensional spectra were extracted using the task \texttt{APALL} by selecting the appropriate aperture, background subtraction, and tracing. The wavelengths were calibrated using the lamp spectra. Flux calibration was performed using the \texttt{standard}, \texttt{sensitivity}, and \texttt{calibration} tasks, incorporating observatory data, exposure times, and standard star information. Subsequently, the \texttt{wspec} task was used to convert the 1-D spectra into a text file, allowing the red and blue parts of the spectra to be merged.
 
The spectrum was then corrected for interstellar extinction using the relation
\begin{equation} 
\label{equ1}
I(\lambda) = F(\lambda) 10^{[c(\mathrm{H}\beta) f(\lambda)]}
\end{equation}
where, $I(\lambda)$ and $F(\lambda)$ represent intrinsic and observed fluxes, respectively. $c(\mathrm{H}\beta)$ is the logarithmic extinction at H$\beta$ and $f(\lambda)$ is the extinction function given by \citet{Cardelli1989}. The value $c(\mathrm{H}\beta)$ is determined by matching the theoretical Balmer line ratios in the de-reddened spectrum. This work adopts the theoretical intrinsic flux ratios, $I(\mathrm{H}\alpha)/I(\mathrm{H}\beta) = 2.847$, $I(\mathrm{H}\gamma)/I(\mathrm{H}\beta) = 0.469$, and $I(\mathrm{H}\delta)/I(\mathrm{H}\beta) = 0.469$,  corresponding to the electron temperature, $T_e = 10^4$ K and electron density, $N_e = 10^4$ cm$^{-3}$ \citep{Osterbrock2006}.The value of $c(\mathrm{H}\beta)$ obtained using H$\alpha$, H$\gamma$, and H$\delta$ as, $c(\mathrm{H}\beta)_{\mathrm{H}\alpha}$, $c(\mathrm{H}\beta)_{\mathrm{H}\gamma}$, and $c(\mathrm{H}\beta)_{\mathrm{H}\delta}$, respectively. The final value of $c(\mathrm{H}\beta)$ is obtained by the following equation,
\begin{equation}
\label{equ2}
c(\mathrm{H}\beta) = \frac{c(\mathrm{H}\beta)_{\mathrm{H}\alpha} \frac{F(\mathrm{H}\alpha)}{F(\mathrm{H}\beta)} + c(\mathrm{H}\beta)_{\mathrm{H}\gamma} \frac{F(\mathrm{H}\gamma)}{F(\mathrm{H}\beta)} + c(\mathrm{H}\beta)_{\mathrm{H}\delta} \frac{F(\mathrm{H}\delta)}{F(\mathrm{H}\beta)}}{\frac{F(\mathrm{H}\alpha)}{F(\mathrm{H}\beta)} + \frac{F(\mathrm{H}\gamma)}{F(\mathrm{H}\beta)} + \frac{F(\mathrm{H}\delta)}{F(\mathrm{H}\beta)}}
\end{equation}
\citep{Bandyopadhyay2021}.
The fluxes for the relevant emission lines were measured using the automated line fitting algorithm, ALFA \citep{Wesson2016}. ALFA subtracts the continuum by sliding a fixed-width window (default 100 data points) across the entire spectrum and computing the 25th percentile flux within each window to estimate the baseline, which remains largely unaffected by emission features; gaps at the spectrum’s edges are filled by extending the nearest percentile values, yielding a smooth, globally consistent continuum that is then subtracted prior to Gaussian fitting of emission lines.

\subsection{Optical band images}
High-resolution images of the planetary nebula NGC~2392, obtained from the Mikulski Archive for Space Telescopes (MAST)\footnote[3]{\url{https://archive.stsci.edu/}} have been used. These observations were carried out with the Hubble Space Telescope using the Wide Field and Planetary Camera~2 (WFPC2) (Proposal ID: 8499, Program type: SM3/ERO)\footnote[4]{Early Release Observations (EROs) for Servicing Mission 3a (SM3a)}. The dataset includes images taken with three narrow-band filters: F502N (with a central wavelength $\lambda_c = 5012\,\text{\AA}$ and a bandwidth $\Delta \lambda = 27\,\text{\AA}$), which maps the [O~III] emission; F656N (with $\lambda_c = 6564\,\text{\AA}$ and $\Delta \lambda = 22\,\text{\AA}$), which traces the H$\alpha$ emission; and F658N (with $\lambda_c = 6591\,\text{\AA}$ and $\Delta \lambda = 29\,\text{\AA}$), which captures the [N~II] emission to construct the composite RGB image of NGC~2392.

Data of the Panoramic Survey Telescope and Rapid Response System (Pan-STARRS, PS1) \citep{Chambers_2016} from the MAST archive are used for NGC 4361. The PS1 system is a 1.8 m aperture $f/4.4$ telescope, which illuminates a 1.4 Gpixel detector stationed at the Haleakala Observatories in Hawaii and is dedicated to sky survey observations. PS1 images in the g($\lambda_{c} = 4810  ~\text{\AA} $) $\mathrm{r}(\lambda_{c} = 6170 ~ \text{\AA} $), and i($\lambda_{c} = 6900 ~ \text{\AA} $) bands \citep{Tonry_2012} are used to construct the composite RGB image of NGC 4361. The summary of the optical images of both objects is given in Table \ref{tab:optical_images}.

\begin{table*} [h!]
\small
\centering
\begin{threeparttable}
\caption{Summary of Optical Band Images\label{tab:optical_images}}
\begin{tabular}{lccccc}
\toprule
\textbf{Instrument} &
\textbf{Filter} &
\textbf{$\lambda_c$ (\AA)} &
\textbf{$\Delta\lambda$ (\AA)} &
\textbf{Exposure (s)} &
\textbf{Observation Date} \\
\midrule
\multicolumn{6}{c}{\textbf{NGC\,2392}} \\
\midrule
HST/WFPC2 & F502N & 5010 & 27 & 200 & 2007 Nov 21 \\
HST/WFPC2 & F656N & 6564 & 22 & 200 & 2007 Nov 21 \\
HST/WFPC2 & F658N & 6584 & 29 & 1200 & 2007 Nov 21 \\
\midrule
\multicolumn{6}{c}{\textbf{NGC\,4361}} \\
\midrule
PS1/GPC1 & g & 4810 & 1370 & 760 & 2010 Mar 21 \\
PS1/GPC1 & r & 6170 & 1390 & 784 & 2010 Feb 23 \\
PS1/GPC1 & i & 6900 & 1290 & 829 & 2010 Feb 25 \\
\bottomrule
\end{tabular}
\end{threeparttable}
\end{table*}

\subsection{Spitzer mid-IR spectra}
We have used the IR spectrum of NGC 2392, which was observed by \cite{Pottasch2008} using the Infrared Spectrograph (IRS; \cite{Houck2004}) on board the Spitzer Space Telescope with AOR keys 4108544 (on target) and 4108800 (background). The observations were made using the short high module (SH) covering 9.9 $\mu m$ to 19.6 $\mu m$ (slit dimension $4.7^{\prime\prime} \times 11.3^{\prime\prime} )$ and long high module (LH) covering 18.7 $\mu m$ to 37.2 $\mu m$. The intensities below 19 $\mu m$ have been increased by a factor of 4.3 to put them on the same scale as the LH intensities measured through a larger diaphragm. 
Table \ref{tab:mid-ir_NGC2392} mentions the identified lines used in this work.

\begin{table}[h!]
\small
\begin{threeparttable}
\caption{Mid-IR and UV Line Intensities of NGC\,2392 $^a$\label{tab:mid-ir_NGC2392}}
\centering
\begin{tabular}{llll}
\toprule
\textbf{Identification} &
\textbf{$\lambda$} &
\textbf{Intensity\textsuperscript{$\dagger$}} & 
\textbf{$I/H\beta$\textsuperscript{$\star$}} \\ 
\midrule

$\mathrm{C~{II}]}$  & 1345 & 11.00  & 16.00 \\
$\mathrm{C~{IV}]}$  & 1548 & 75.50  & 110.00 \\
$\mathrm{C~{III}]}$ & 1909 & 148.00 & 216.00 \\
\midrule
$\mathrm{[Ar~{II}]}$ & 6.989 & 18.30 $\pm$ 3.50 & 0.64 \\
$\mathrm{[Ar~{III}]}$ & 8.995 & 231.00 $\pm$ 8.80 & 8.05 \\
$\mathrm{[S~{IV}]}$ & 10.511 & 950.00 $\pm$ 56.00 & 33.0 \\
$\mathrm{[Cl~{IV}]}$ & 11.762 & 4.69 $\pm$ 0.60 & 0.16 \\
$\mathrm{[Ne~{II}]}$ & 12.815 & 220.00 $\pm$ 10.00 & 8.63 \\
$\mathrm{[Ne~{V}]}$ & 14.323 & 44.60 $\pm$ 0.86 & 1.56 \\
$\mathrm{[Ne~{III}]}$ & 15.556 & 1930.00 $\pm$ 30.8 & 67.20 \\
$\mathrm{[Fe~{II}]}$ & 17.943 & 10.00 $\pm$ 40.00 & 0.35 \\
$\mathrm{[Fe~{III}]}$ & 18.718 & 8.97 $\pm$ 0.82 & 0.31 \\
$\mathrm{[S~{III}]}$ & 18.718 & 1098.00 $\pm$ 27.00 & 38.20 \\
$\mathrm{[Cl~{IV}]}$ & 20.319 & 4.91 $\pm$ 0.36 & 0.16 \\
$\mathrm{[Ar~{III}]}$ & 21.827 & 25.60 $\pm$ 1.96 & 0.89 \\
$\mathrm{[Fe~{III}]}$ & 22.934 & 44.70 $\pm$ 1.37 & 1.55 \\
$\mathrm{[Ne~{V}]}$ & 24.326 & 44.50 $\pm$ 3.25 & 1.55 \\
$\mathrm{[O~{IV}]}$ & 25.895 & 2634.00 $\pm$ 119.00 & 88.40 \\
$\mathrm{[Fe~{III}]}$ & 33.043 & 14.70 $\pm$ 1.31 & 0.51 \\
$\mathrm{[S~{III}]}$ & 33.487 & 530.00 $\pm$ 11.50 & 18.40 \\
$\mathrm{[Si~{II}]}$ & 34.825 & 45.40 $\pm$ 1.69 & 1.58 \\
$\mathrm{[Ne~{III}]}$ & 35.941 & 97.70 $\pm$ 115.00 & 3.42 \\
\bottomrule
\end{tabular}
\textsuperscript{a} \cite{Pottasch2008}\\
\textsuperscript{$\dagger$} Intensities measured in units of $10^{-14} \mathrm{erg~cm^{-2}~s^{-1}}$ \\
\textsuperscript{$\star$} Extinction corrected $H\beta = 6.83 \times 10^{-11} \mathrm{erg~cm^{-2}~s^{-1}}$\\
$\lambda$ is in $\AA$, for Carbon and in $\mu$m for other elements.
\end{threeparttable}
\end{table}

\section{Results }

\subsection{ NGC 2392 (PN G197.8+17.3)}
\begin{figure}[h!]
    \centering
    \includegraphics[width=1\linewidth]{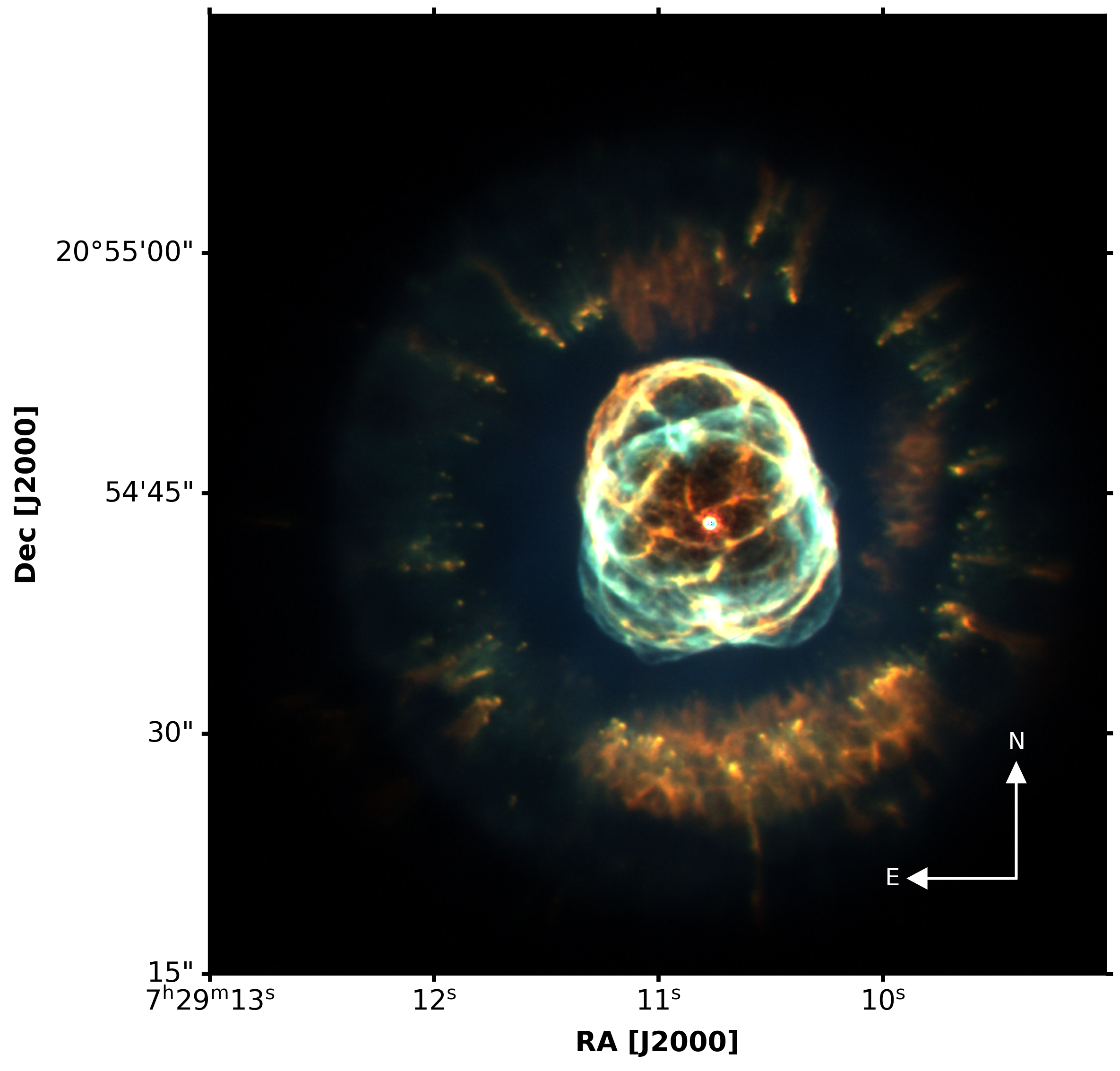}
    \caption{A composite-color image of NGC 2392, here N II (F658N) is represented as red, H$\alpha$ (F656N) is represented as green, and O III (F502N) is represented as blue.}
    \label{fig:RGB_NGC2392}
\end{figure}

\begin{figure*}[h!]
    \centering
    \includegraphics[width=0.845\linewidth]{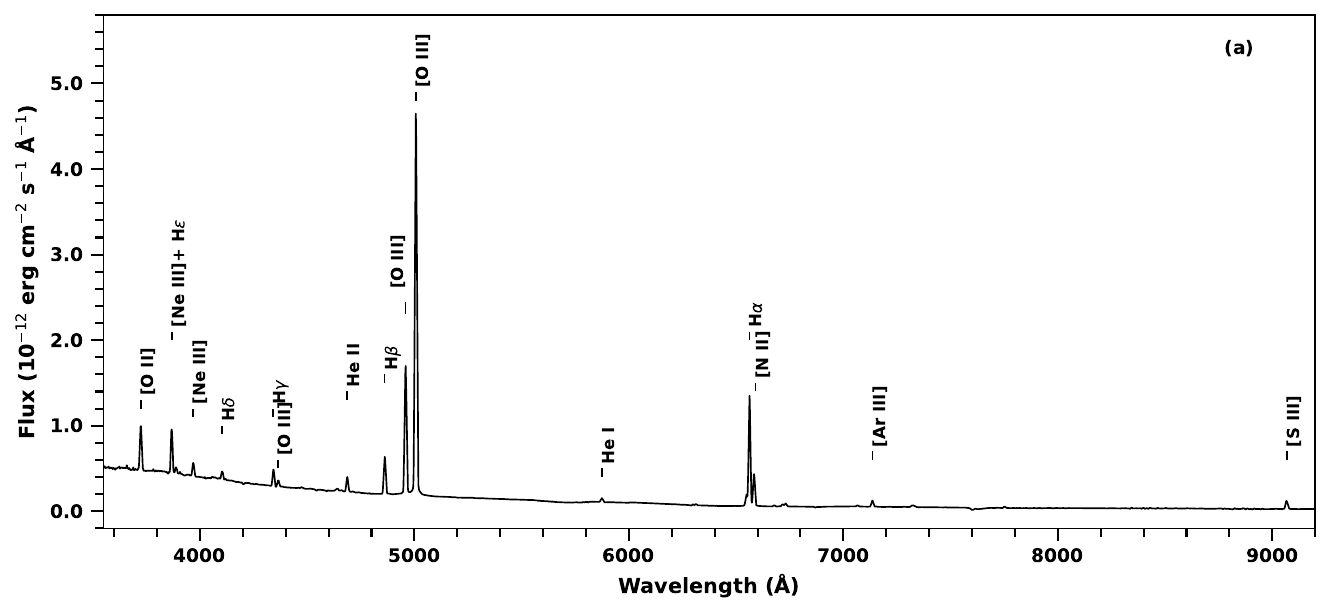}
    \includegraphics[width=0.845\linewidth]{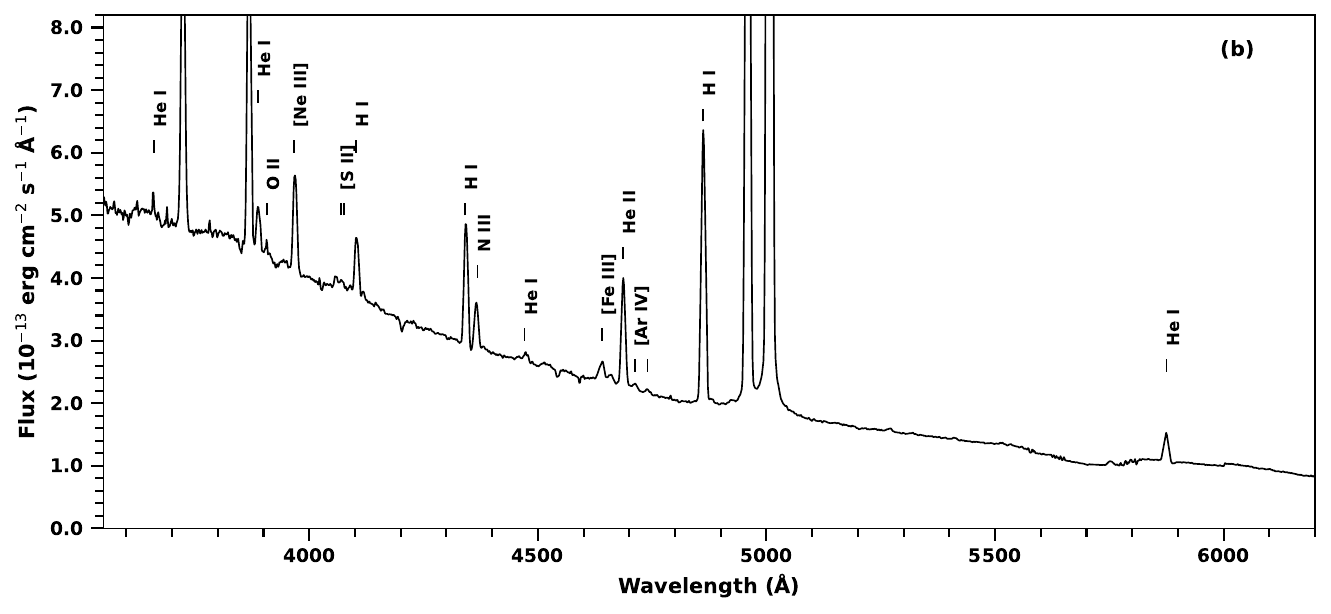}
    \includegraphics[width=0.845\linewidth]{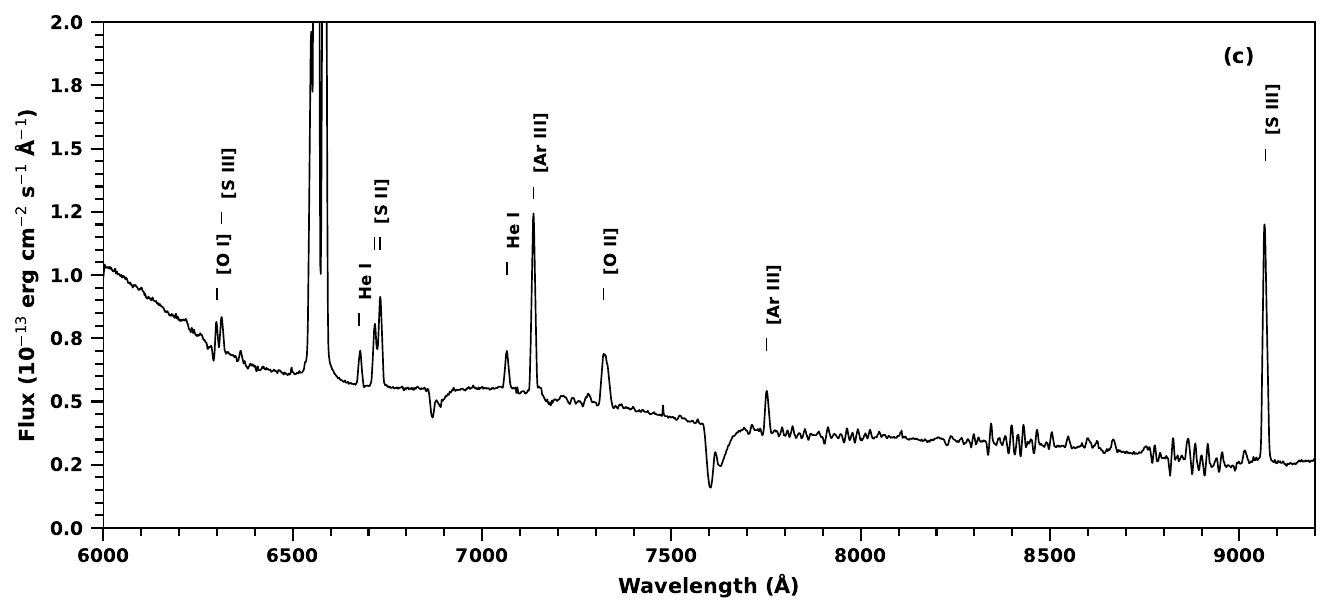}
    \caption{The HCT optical spectrum of the planetary nebula NGC 2392, calibrated in intrinsic absolute fluxes, is displayed across three panels. Panel (a) presents the full spectrum, dominated by its brightest emission lines. Panels (b) and (c) zoom into the fainter features, with (b) focusing on the shorter-wavelength (blue) region and (c) highlighting the longer-wavelength (red) region of the spectrum.}
    \label{fig:Spectra_2392}
\end{figure*}

\subsubsection{Interstellar extinction coefficient:}
The effect of interstellar extinction on NGC 2392 is relatively small, and there is evidence that it is inhomogeneous over the nebular surface \citep{Zipoy1976, Barker1991}. From $c{(H\beta)}_{H\alpha}$ = 0.119, $c{(H\beta)}_{H\gamma}$ = 0.477, and $c{(H\beta)}_{H\delta}$ = 0.179,   we obtain $c_{H\beta}$ = 0.168, in agreement with the value 0.13 derived by \citet{Barker1991}. \citet{Liu1993} and \citet{Zhang2012} reported slightly higher values, 0.22 and 0.27, respectively. Their higher values may be due to long-slit spectra that samples the dense regions where the extinction by the local dust grains is large. Figure \ref{fig:RGB_NGC2392} shows a composite-color image of NGC 2392 in optical bands. 

\subsubsection{Spectral features:}  
 Figure \ref{fig:Spectra_2392} shows the optical spectra of NGC 2392. A total of 53 atomic lines belonging to 17 species are detected. The list of detected lines in the spectrum includes the commonly observed strong recombination lines (RL), such as the Hydrogen Balmer lines $H\alpha \;6563, H\beta \;4861, H\gamma \;4340, H\delta \;4101, H\epsilon \;3970 $ \text{\AA} and He I 4471, 5876, 6678, and 7065 \text{\AA} lines, He II 4686, 5411 \text{\AA} lines. The C II lines are not observed with confidence in our spectrum. Using the flux value of He II (4686 \AA) and $H\beta$ (4861 \AA), we calculated the excitation class (EC)\footnote[5]
 {The \textit{Excitation Class} (EC) characterizes the ionization level and energy distribution of radiation emitted by the central star in PNe. It is defined by comparing nebular emission-line fluxes, typically He~II~($\lambda4686$) relative to H$\beta$~($\lambda4861$). Higher EC values correspond to nebulae ionized by hotter central stars emitting more energetic photons capable of ionizing helium and heavier elements \citep{Morgan1984, Dopita1990}.} = 6.72 using the expression derived by \citet{Dopita1990} from the scheme given by \citet{Morgan1984}. This value of EC ($5<EC<7$) suggests that this nebula is a medium to high-excitation nebula. PNe excitation levels are classified based on their ionization states, reflected in emission-line ratios (e.g., He~II/H$\beta$). A \textit{high-excitation nebula} indicates the presence of highly energetic photons emitted by a hotter central star, sufficient to ionize helium and heavier elements, leading to strong emission from ions such as He~II and [O~III]. Conversely, a \textit{low-excitation nebula} is primarily ionized by lower-energy photons from cooler central stars, showing weaker or absent emission from highly ionized species \citep{Morgan1984, Dopita1990}.
This classification is also supported by its spectral features, particularly the dominance of [O III] emission lines, which indicate high-excitation conditions. Among the collisionally excited lines (CELs), we have prominently detected [N II] 5755, 6548, 6583 \text{\AA}; [O III] 5007, 4959 \text{\AA} and the auroral [O III] 4363; [S III] 6312, 9069 \text{\AA}; [Ne III] 3869 \text{\AA}; and the doublets [S II] 6716, 6731 \text{\AA};  [Ar IV] 4711, 4740 \text{\AA}; [Cl III] 5518, 5538 \text{\AA} lines. The observed lines are listed in Table \ref{tab:opt_lines_NGC2392}.

\subsubsection{Diagnostics and Ionic Concentrations:}
To derive the nebula's physical conditions and chemical abundances, we employed \texttt{PyNeb} \citep{2015_Luridiana}, a Python-based tool designed for analyzing ionized nebular spectra. We used this package to derive the electron temperature \(T_e\) and electron density \(n_e\) from diagnostic emission-line ratios. These values were then used to calculate the ionic abundances, with each ion analyzed using the temperature and density appropriate to its ionization potential. Elemental abundances were subsequently estimated by summing the available ionic abundances, where a sufficient number of ionization stages were detected.

The density and temperature can be obtained simultaneously using the \texttt{getCrossTemDen} function in \texttt{PyNeb}. This method iteratively solves the level population equations using precomputed emissivity grids derived from atomic data (CHIANTI) \citep{Dere2019}, ensuring consistency between \(T_e\) and \(n_e\). The observed line ratios are compared with theoretical values, and discrepancies are minimized through interpolation and iterative adjustments. This diagnostic approach has been widely applied in previous studies of planetary nebulae and H II regions \citep[e.g.,][]{GarciaRojas2020, DelgadoInglada2014} and provides reliable physical conditions for subsequent abundance determinations. A comparison of the [N III] 5755/6548 nebular-type transition with [S II] 6731/6716 gives \(T_e\) = 10545 K and \(n_e\) = 2900 $cm^{-3}$, while a comparison of [O III] 4363/6716 with [S II] gives  \(T_e\) = 11150 K and \(n_e\) = 2958 $cm^{-3}$. Thus, rounding off we use \(T_e\) = 11000 K and \(n_e\) = 3000 $cm^{-3}$ to calculate the ionic and elemental abundances. It is noted that abundance changes significantly when \(n_e\) $>$ 3000 $cm^{-3}$.  

\begin{figure}[h!]
    \centering
    \includegraphics[width=1\linewidth]{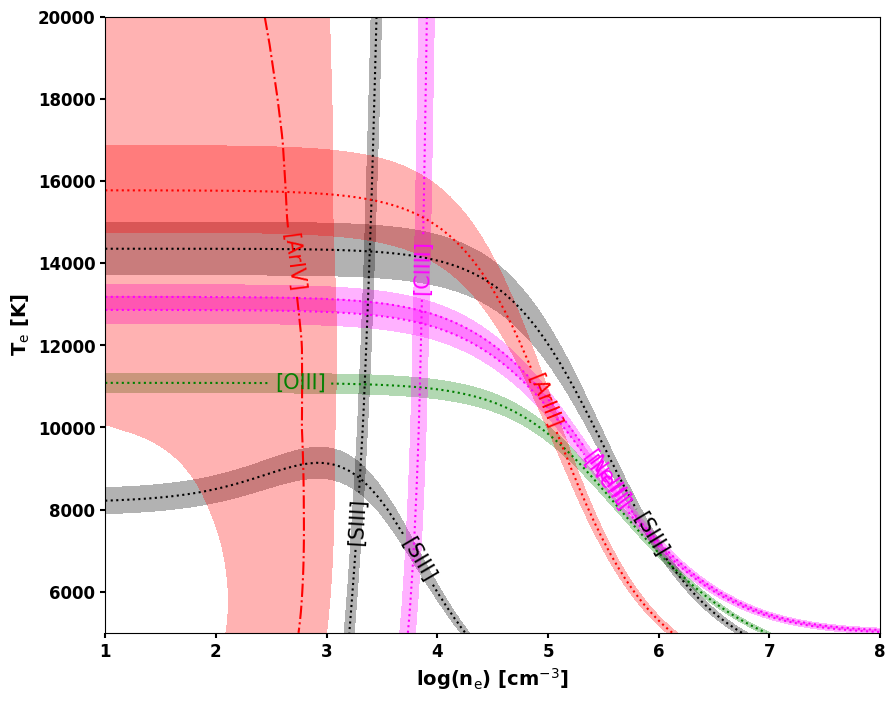}
    \caption{Diagnostic diagram of NGC 2392. The labels on the graph denote the corresponding ions. Colored width represents the one sigma rms error of each diagnostic.}
    \label{fig: Diagonostic_2392}
\end{figure}

We calculated the \(T_e\) and \(n_e\) with the \texttt{getTemDen} function using the known diagnostic emission line flux ratios ([O III], [S III], [Ne III], and [Ar III] for calculating \(T_e\) and [Ar IV], [Cl III], [S II], [Ne III], and [S III] for \(n_e\) calculation). The \(T_e\) and \(n_e\) values corresponding to the line flux ratio are listed in Table \ref{tab:tem_den_NGC2392} and the diagnostic diagram is shown in Figure \ref{fig: Diagonostic_2392}.

\begin{table}[h!]
\begin{threeparttable}
\caption{Temperatures and Densities in NGC 2392\label{tab:tem_den_NGC2392}}
\centering
\small
\begin{tabular}{llll}
\toprule
\textbf{Diagnostic Line ID} & \textbf{Ratio} & \textbf{Quantity} & \textbf{Value$^\star$} \\
\midrule
{[O~\textsc{iii}]} 4363/5007+4959 & 0.0065 & $T_e$ & 10990 \\
{[S~\textsc{iii}]} 6312/9069 & 0.144 & $T_e$ & 14970 \\
{[Ne~\textsc{iii}]} 3869/15.6$\mu$m & 1.239 & $T_e$ & 14250 \\
{[Ar~\textsc{iii}]} 7136/9.0$\mu$m & 1.925 & $T_e$ & 18000 \\
{[S~\textsc{iii}]} 9069/18.7$\mu$m & 0.540 & $T_e$ & 8900 \\
\midrule
{[Ar~\textsc{iv}]} 4740/4711 & 0.756 & $n_e$ & 600 \\
{[Cl~\textsc{iii}]} 5538/5518 & 1.534 & $n_e$ & 6951 \\
{[S~\textsc{ii}]} 6731/6716 & 1.473 & $n_e$ & 2940 \\
{[S~\textsc{iii}]} 18.7$\mu$m/33.5$\mu$m & 2.070 & $n_e$ & 2200 \\
\bottomrule
\end{tabular}
$^\star$ Electron temperatures ($T_e$) are in K, and electron densities ($n_e$) are in cm$^{-3}$.
\end{threeparttable}
\end{table}
 
All potential ionic abundances for our dataset have been calculated. \texttt{PyNeb} has been employed to compute ionic abundances using the formula $\left(\frac{(\mathrm{X_i})}{(\mathrm{H^+})}\right)$, where $(\mathrm{X_i})$ denotes the density of the ion for which the abundance is determined and $(\mathrm{H^+})$ signifies the ionic hydrogen density. The obtained ionic abundances are detailed in Table \ref{ionic_abun_NGC2392}.

\begin{table}
\centering
\begin{threeparttable}
\caption{Ionic Abundances in NGC 2392\label{ionic_abun_NGC2392}}
\small
\begin{tabular}{lll}
\toprule
\textbf{Ion} & \textbf{Lines$^\star$} & \textbf{X$^{+}$/H$^{+}$} \\
\midrule
He$^{+}$    & He~\textsc{i}  $\lambda$4471       & $4.735 \times 10^{-2}$ \\
            & He~\textsc{i}  $\lambda$5876       & $7.401 \times 10^{-2}$ \\
            & He~\textsc{i}  $\lambda$7065       & $7.740 \times 10^{-2}$ \\
He$^{2+}$   & He~\textsc{ii} $\lambda$4686       & $3.364 \times 10^{-2}$ \\
            & He~\textsc{ii} $\lambda$5411       & $1.950 \times 10^{-2}$ \\
\midrule
C$^{+}$    & C~\textsc{ii}]  $\lambda$1345   & $ 2.262 \times 10^{-4}$ \\
C$^{2+}$   & C~\textsc{iii}]  $\lambda$1548  & $ 2.747\times 10^{-4}$ \\
C$^{3+}$   & C~\textsc{iv}]  $\lambda$1909   & $ 1.864\times 10^{-4}$ \\

\midrule
N$^{+}$     & [N~\textsc{ii}] $\lambda$5754      & $1.547 \times 10^{-5}$ \\
            & [N~\textsc{ii}] $\lambda$6548      & $1.682 \times 10^{-5}$ \\
            & [N~\textsc{ii}] $\lambda$6584      & $1.950 \times 10^{-5}$ \\
\midrule
O$^{0}$     & [O~\textsc{i}]   $\lambda$6300     & $3.199 \times 10^{-6}$ \\
            & [O~\textsc{i}]   $\lambda$6364     & $7.087 \times 10^{-6}$ \\
O$^{+}$     & [O~\textsc{ii}]  $\lambda$3726     & $7.475 \times 10^{-5}$ \\
O$^{2+}$    & [O~\textsc{iii}] $\lambda$4363     & $4.788 \times 10^{-4}$ \\
            & [O~\textsc{iii}] $\lambda$4959     & $3.498 \times 10^{-4}$ \\
            & [O~\textsc{iii}] $\lambda$5007     & $3.658 \times 10^{-4}$ \\
O$^{3+}$    & [O~\textsc{iv}]  $\lambda$25.9     & $7.216 \times 10^{-5}$ \\
\midrule
Ne$^{+}$    & [Ne~\textsc{ii}]  $\lambda$12.8    & $1.480 \times 10^{-5}$ \\
Ne$^{2+}$   & [Ne~\textsc{iii}] $\lambda$15.6    & $6.089 \times 10^{-5}$ \\
            & [Ne~\textsc{iii}] $\lambda$36.0    & $4.144 \times 10^{-5}$ \\
            & [Ne~\textsc{iii}] $\lambda$3869    & $1.227 \times 10^{-4}$ \\
Ne$^{4+}$   & [Ne~\textsc{v}]   $\lambda$14.3    & $1.886 \times 10^{-7}$ \\
            & [Ne~\textsc{v}]   $\lambda$24.3    & $5.004 \times 10^{-7}$ \\
\midrule
Ar$^{+}$    & [Ar~\textsc{ii}]  $\lambda$7.0     & $7.102 \times 10^{-8}$ \\
Ar$^{2+}$   & [Ar~\textsc{iii}] $\lambda$9.0     & $8.180 \times 10^{-7}$ \\
            & [Ar~\textsc{iii}] $\lambda$21.8    & $1.117 \times 10^{-6}$ \\
            & [Ar~\textsc{iii}] $\lambda$5192    & $1.403 \times 10^{-5}$ \\
            & [Ar~\textsc{iii}] $\lambda$7136    & $1.422 \times 10^{-6}$ \\
            & [Ar~\textsc{iii}] $\lambda$7751    & $1.432 \times 10^{-6}$ \\
Ar$^{3+}$   & [Ar~\textsc{iv}]  $\lambda$4711    & $1.825 \times 10^{-6}$ \\
            & [Ar~\textsc{iv}]  $\lambda$4740    & $9.075 \times 10^{-7}$ \\
Ar$^{4+}$   & [Ar~\textsc{v}]   $\lambda$7005    & $1.237 \times 10^{-8}$ \\
\midrule
S$^{+}$     & [S~\textsc{ii}]   $\lambda$4069    & $6.619 \times 10^{-7}$ \\
            & [S~\textsc{ii}]   $\lambda$4076    & $1.808 \times 10^{-6}$ \\
            & [S~\textsc{ii}]   $\lambda$6716    & $1.026 \times 10^{-6}$ \\
            & [S~\textsc{ii}]   $\lambda$6731    & $7.874 \times 10^{-7}$ \\
S$^{2+}$    & [S~\textsc{iii}]  $\lambda$18.7    & $8.169 \times 10^{-6}$ \\
            & [S~\textsc{iii}]  $\lambda$3722    & $2.740 \times 10^{-4}$ \\
            & [S~\textsc{iii}]  $\lambda$6312    & $6.814 \times 10^{-6}$ \\
            & [S~\textsc{iii}]  $\lambda$9069    & $3.541 \times 10^{-6}$ \\
S$^{3+}$    & [S~\textsc{iv}]   $\lambda$10.5    & $1.695 \times 10^{-6}$ \\
\midrule
Si$^{+}$    & [Si~\textsc{ii}]  $\lambda$34.8    & $3.246 \times 10^{-6}$ \\
\midrule
Cl$^{2+}$   & [Cl~\textsc{iii}] $\lambda$5518    & $3.468 \times 10^{-7}$ \\
            & [Cl~\textsc{iii}] $\lambda$5538    & $2.915 \times 10^{-7}$ \\
Cl$^{3+}$   & [Cl~\textsc{iv}]  $\lambda$11.8    & $1.376 \times 10^{-8}$ \\
            & [Cl~\textsc{iv}]  $\lambda$20.3    & $2.888 \times 10^{-8}$ \\
\midrule
Fe$^{2+}$   & [Fe~\textsc{iii}] $\lambda$4659    & $1.925 \times 10^{-6}$ \\
            & [Fe~\textsc{iii}] $\lambda$4755    & $3.430 \times 10^{-6}$ \\
\bottomrule
\end{tabular}
$^\star$ Wavelength units are in \AA\ when four-digit numbers are given, and in $\mu$m for longer wavelengths (i.e., IR lines).
\end{threeparttable}
\end{table}

\subsubsection{Total Elemental Abundances:} Total abundances of He, C, O, N, Ne, Ar, S, Fe, Si, and Cl, relative to H are computed from the measured ionic abundances ($\mathrm{X}^{i+}/\mathrm{H}^+$) by applying Ionization Correction Factors \footnote[6] {Ionization Correction Factors (ICFs) are used to estimate the total elemental abundances in PNe when not all ionization stages of an element are observable. These factors are based on photoionization models or empirical correlations and are essential for deriving accurate total abundances from incomplete ionic data \citep{Kingsburgh1994}.} (ICFs). The elemental abundances of Ar, C, He, Ne, O, and S were obtained using direct sums of all significantly detected ions: Ar ({Ar}$^{2+}$, {Ar}$^{3+}$, {Ar}$^{4+}$, {Ar}$^{5+}$); C ({C}$^{+}$, {C}$^{2+}$, {C}$^{3+}$); He ({He}$^{+}$ and {He}$^{2+}$ derived from He\,\textsc{i} recombination lines); Ne ({Ne}$^{2+}$, {Ne}$^{3+}$, {Ne}$^{5+}$); O ({O}$^{2+}$, {O}$^{3+}$, {O}$^{4+}$); and S ({S}$^{2+}$, {S}$^{3+}$, {S}$^{4+}$). The abundance of Fe was computed using the ICF provided by \citet{Rodriguez2005}, relying on the presence of {O}$^{2+}$ and {O}$^{3+}$ ions. Nitrogen abundances were calculated using the ICF from \citet{TorresPeimbert1977}, based on the ratio ({O}$^{2+}$+{O}$^{3+})/${O}$^{2+}$. Chlorine abundances employed the ICF scheme from \citet{Kwitter2001} derived from helium ionic abundances.
The derived abundance values, presented in units of $12 + \log(\mathrm{X}/\mathrm{H})$, are listed in Table~\ref{tab:elemental_abud_NGC2392} and compared with earlier reports and the solar abundance.
Our value for oxygen in NGC 2392 ($12+log(O/H) = 8.75$) falls within the broader range reported in the literature (e.g., \cite{Pottasch2008} 8.46; \cite{Henry2000}: 8.45), though it is on the higher side. Our analysis uses a combination of optical and infrared lines, with updated atomic data and cross-validated physical conditions. Differences may arise from the spatial extraction (sampling denser, higher-excitation regions), improved flux calibration, or adoption of recent atomic coefficients. 

\begin{table*}[h!]
\centering
\begin{threeparttable}
\caption{Elemental Abundances in NGC 2392, presented in units of $12 + \log(\mathrm{X}/\mathrm{H})$\label{tab:elemental_abud_NGC2392}}
\small 
\begin{tabular}{lcccccc}
\toprule
\textbf{Element} & \textbf{This Work} &
\textbf{Zhang\textsuperscript{a}} &
\textbf{Pottasch\textsuperscript{b}} &
\textbf{Henry\textsuperscript{c}} &
\textbf{Barker\textsuperscript{d}} &
\textbf{Solar\textsuperscript{e}} \\
\midrule
He & 10.62 & 10.99 & 10.90 & 10.88 & 10.99 & 10.90 \\
C & 8.31 & --- & 8.52 & 8.34 & 7.62 & 8.39 \\
N & 7.88 & 7.50 & 8.27 & 8.04 & 8.04 & 7.83 \\
O & 8.75 & 8.54 & 8.46 & 8.45 & 8.53 & 8.69 \\
Ne & 7.16 & 8.07 & 7.92 & 7.81 & 7.88 & 8.05 \\
S & 6.69 & 6.37 & 6.70 & --- & 6.63 & 7.19 \\
Cl & 4.78 & 5.12 & 5.11 & --- & --- & 5.50 \\
Ar & 5.20 & 6.25 & 6.34 & --- & 6.15 & 6.55 \\
Fe & 6.18 & 6.53 & 5.90 & --- & --- & 7.47 \\
Si & 6.84 & --- & --- & --- & --- & --- \\
\midrule
C/O & 0.36 & --- & 0.72 & 0.62 & 0.12 & 0.49 \\
N/O & 0.14 & 0.06 & 0.65 & 0.39 & 0.32 & 0.14 \\
\bottomrule
\end{tabular}
\textsuperscript{a}\citet{Zhang2012} \\
\textsuperscript{b}\citet{Pottasch2008} \\
\textsuperscript{c}\citet{Henry2000} \\
\textsuperscript{d}\citet{Barker1991} \\
\textsuperscript{e}\citet{Lodders2003}
\end{threeparttable}
\end{table*}

\subsection{NGC 4361 (PN G294.1 +43.6)} 
\begin{figure}[h!]
    \centering
    \includegraphics[width=1\linewidth]{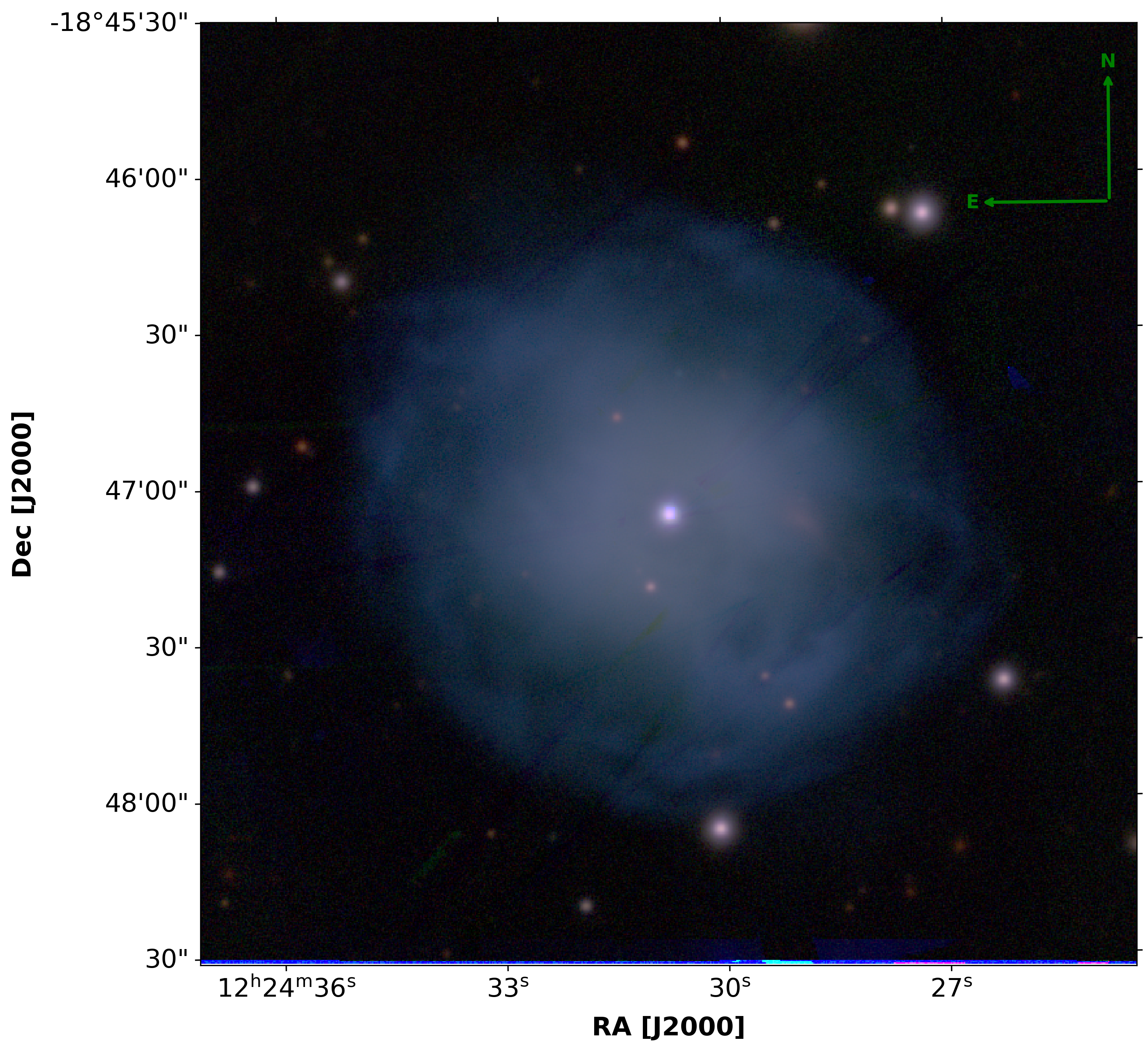}
    \caption{Composite-color image of NGC 4361. i band is represented as red, the r band is represented as green, and the g band frame is represented as blue. }
    \label{fig: RGB_NGC4361}
\end{figure}

\subsubsection{Interstellar extinction coefficient:}

In NGC~4361, several Balmer lines, most notably H$\beta$, H$\gamma$, and H$\delta$, are blended with He\,\textsc{ii} Pickering transitions at $\lambda$4859.32, $\lambda$4338.67, and $\lambda$4100.04, respectively. This contamination alters the observed line ratios, leading to anomalous values when compared to standard Case~B predictions. For example, the observed F(H$\alpha$)/F(H$\beta$) ratio is lower than the theoretical value of 2.847, resulting in a negative value of $c(\mathrm{H}\beta)$ when using H$\alpha$ for extinction estimation. 

To address this, we corrected the observed H$\gamma$ and H$\delta$ fluxes by subtracting the estimated contribution from nearby He\,\textsc{ii} Pickering lines. This correction was guided by the strength of unblended He\,\textsc{ii} $\lambda$4686 and applied following prescriptions from \citet{Liu2006}. The corrected Balmer ratios were then compared with theoretical values computed for $T_e = 18{,}000$~K and $n_e = 1200$~cm$^{-3}$ using the Galactic reddening law of \citet{Howarth1983}. This yielded a consistent extinction coefficient of $c(\mathrm{H}\beta) = 0.07$, which we adopted for de-reddening the spectrum. This value also agrees with previous estimates (e.g., \citealt{Torres1990}; \citealt{Walsh2024}), validating our approach in this high-excitation, He\,\textsc{ii}-rich nebula. A composite-color image of NGC 4361 in i, r, and g bands is shown in Figure \ref{fig: RGB_NGC4361}.

\begin{figure*}[h!]
    \centering
    \includegraphics[width=0.845\linewidth]{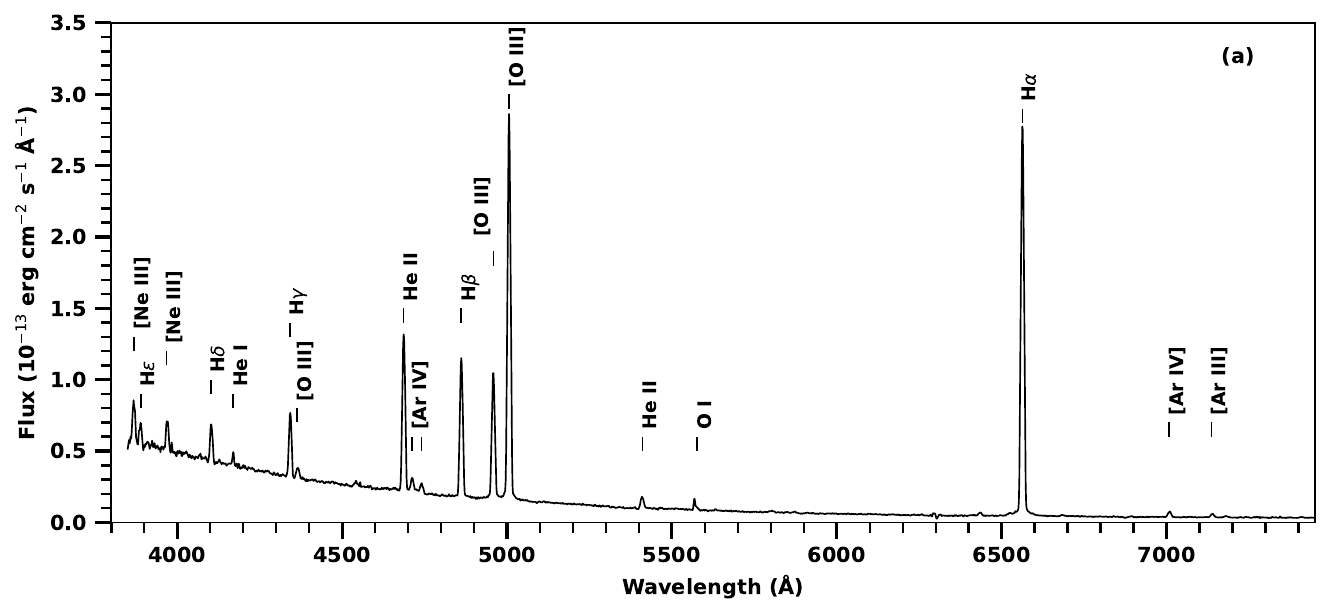}
    \includegraphics[width=0.845\linewidth]{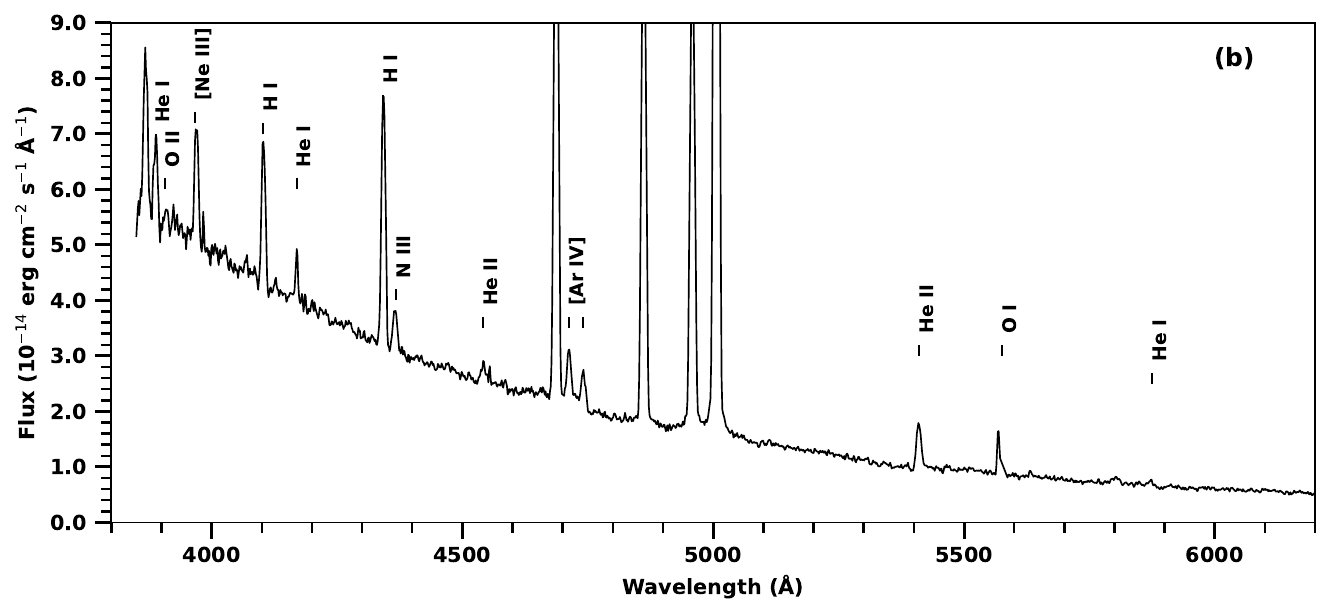}
    \includegraphics[width=0.845\linewidth]{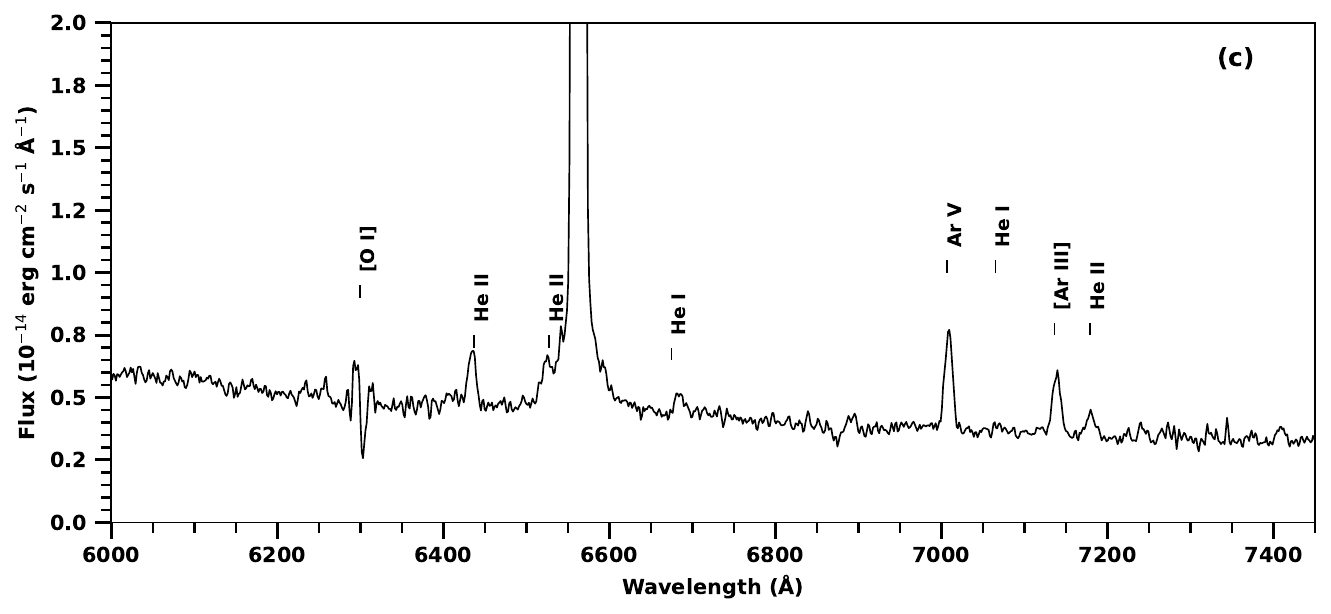}
    \caption{The HCT optical spectrum of the planetary nebula NGC 4361, calibrated in intrinsic absolute fluxes, is displayed across three panels. Panel (a) presents the full spectrum, dominated by its brightest emission lines. Panels (b) and (c) zoom into the fainter features, with (b) focusing on the shorter-wavelength (blue) region and (c) highlighting the longer-wavelength (red) region of the spectrum.}
    \label{fig:Spectra_4361}
\end{figure*}

\subsubsection{Spectral Features:}
Figure \ref{fig:Spectra_4361} displays the optical spectra of NGC 4361. The intrinsic absolute fluxes of detected emission lines are listed in Table \ref{tab:linelist_NGC4361}. 70 atomic lines are identified that belong to 26 species. Prominent among these are the strong recombination lines (RLs) such as $H\alpha;6563, H\beta;4861$ $ H\gamma;4340, H\delta;4101, H\epsilon;3970$ \text{\AA} and He I lines at 4471, 5876, 6678, and 7065 \text{\AA}, along with He II lines at 4686 and 5411 \text{\AA}. The intense emission of He II at 4685 \text{\AA} has a significantly high value, indicating an exceptionally high excitation class. Utilizing the flux values of He II (4685 \text{\AA}) and $H\beta$, the calculated EC exceeds 10, placing NGC 4361 at the maximum of the nebular ionization scale \citep{Dopita1990}. This high EC suggests that NGC 4361 is optically thin in the Lyman continuum, characteristic of a matter-bounded nebula. The high ionization is also reported in the earlier reports (\cite{Heap1969}; \cite{Aller1978}; \cite{Torres1990}).CELs such as [O III] at 5007 and 4959 \text{\AA} are strong, aligning with the nebula's high ionization state.

\subsubsection{Diagnostic and Ionic Concentration:}
\begin{figure}[h!]
    \centering
    \includegraphics[width=1\linewidth]{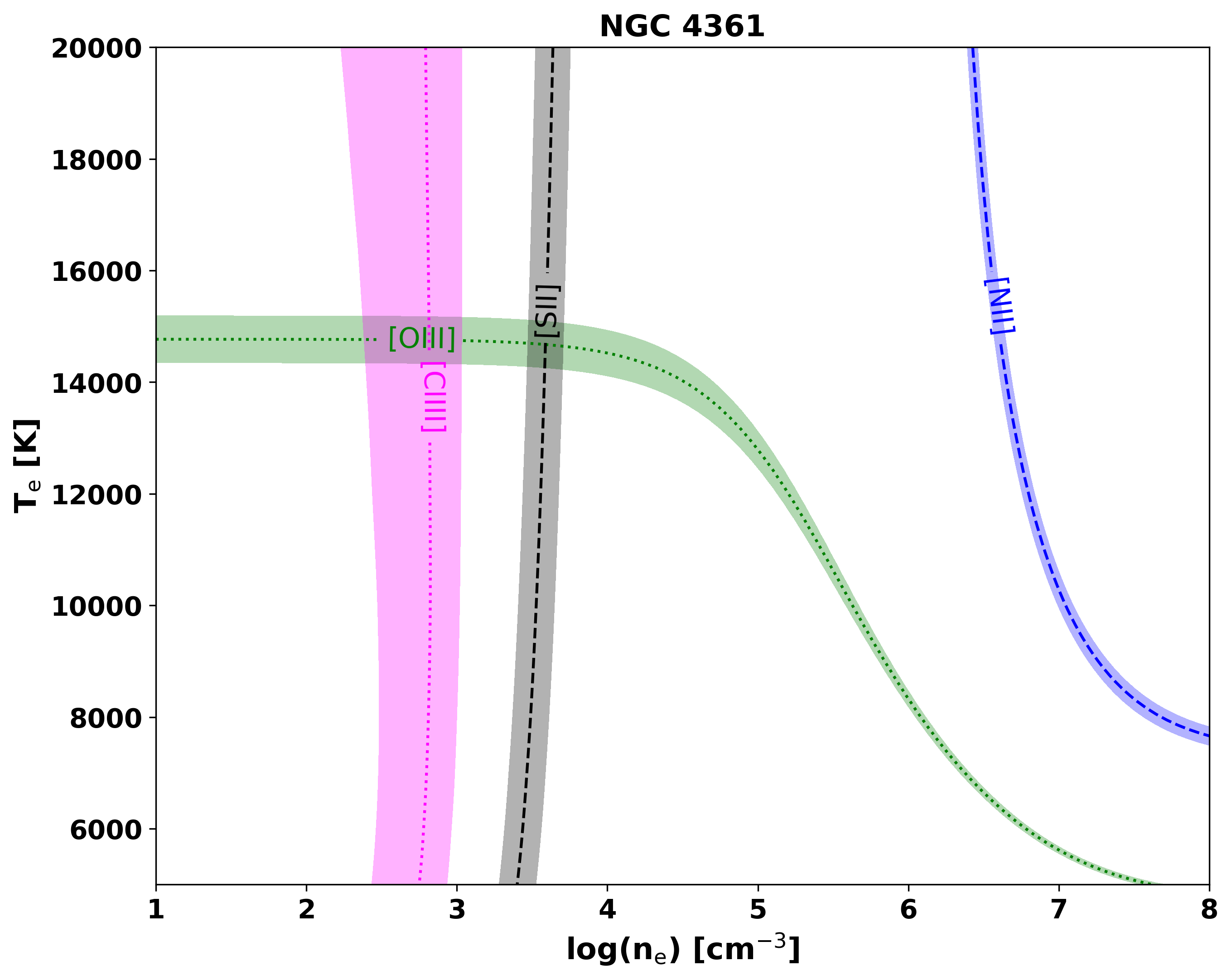}
    \caption{Diagnostic diagram of NGC 4361. The labels on the graph denote the corresponding ions. Colored width represent the 1$\sigma$ rms error of each diagnostic.}
    \label{fig:diago_4361}
\end{figure}

Following the methods outlined earlier in this paper, we analyzed the ionized nebular spectra of NGC 4361 using the Python-based tool PyNeb. The process to determine $T_e$ and $n_e$ involved using PyNeb's \texttt{getCrossTemDen} function. This method ensures accurate measurement by iteratively adjusting the values to match the observed spectral line ratios with those predicted theoretically.

Initially, the attempt to derive $T_e$ and $n_e$ values using the line ratio [N II] 5755/6548 paired with [S II] 6731/6716 did not converge because PyNeb uses interpolation over precomputed atomic data grids, and input ratio combination implies $T_e$ or $n_e$ values outside the supported range. However, further analysis using [O III] 4363/5007 combined with [S II] 6731/6716 successfully provided us with $T_e = 14570$ K and $N_e = 3856$ cm$^{-3}$. These values suggest a high ionization level within NGC 4361, indicative of its energetic and unique environmental conditions. Table \ref{tab:tem_den_NGC4361} details the temperatures and densities derived from various diagnostic lines for NGC 4361, and the diagnostic diagram is shown in Figure \ref{fig:diago_4361}. Ionic abundances of NGC 4361 are obtained relative to H and shown in Table~\ref{tab:ionic_ngc4361}, electron temperature $T_e = 14500$~K and electron density $n_e = 3850$~cm$^{-3}$) is used.

\begin{table}[h!]
\centering
\begin{threeparttable}
\caption{Temperatures and Densities for NGC 4361\label{tab:tem_den_NGC4361}}
\small
\begin{tabular}{llll}
\toprule
\textbf{Diagnostic Line ID} & \textbf{Ratio} & \textbf{Quantity} & \textbf{Value$^\dagger$} \\
\midrule
{[O~\textsc{iii}]} 4363/5007+4959 & 0.014 & $T_e$ & 14500 \\
\midrule
{[Ar~\textsc{iv}]} 4740/4711 & 1.5 & $n_e$ & 10065 \\
{[Cl~\textsc{iii}]} 5538/5518 & 0.8 & $n_e$ & 4675 \\
{[S~\textsc{ii}]} 6731/6716 & 1.5 & $n_e$ & 3482 \\
{[N~\textsc{ii}]} 5755/6548 & 8.7 & $n_e$ & 38058\textsuperscript{*} \\
\bottomrule
\end{tabular}
$^\dagger$ Electron temperature ($T_e$) is in K, and electron densities ($n_e$) are in cm$^{-3}$.\\
\textsuperscript{* }Value is uncertain due to blending of $\lambda$6548 with H$\alpha$.

\end{threeparttable}
\end{table}

\begin{table}[h!]
\centering
\begin{threeparttable}
\caption{Ionic Abundances of NGC 4361}
\label{tab:ionic_ngc4361}
\small 
\begin{tabular}{lll}
\toprule
\textbf{Ion} & \textbf{Lines (\AA)} & \textbf{X$^{+}$/H$^{+}$} \\
\midrule
He$^{+}$    & He~\textsc{i} $\lambda$4471        & $8.512 \times 10^{-1}$ \\
            & He~\textsc{i} $\lambda$5876        & $8.221 \times 10^{-1}$ \\
            & He~\textsc{i} $\lambda$7065        & $7.963 \times 10^{-1}$ \\
He$^{2+}$   & He~\textsc{ii} $\lambda$4541       & $1.044 \times 10^{-1}$ \\
            & He~\textsc{ii} $\lambda$4686       & $9.087 \times 10^{-2}$ \\
            & He~\textsc{ii} $\lambda$5411       & $2.008 \times 10^{-2}$ \\
\midrule
N$^{+}$     & [N~\textsc{ii}] $\lambda$5754      & $1.662 \times 10^{-5}$ \\
            & [N~\textsc{ii}] $\lambda$6548      & $1.078 \times 10^{-7}$ \\
            & [N~\textsc{ii}] $\lambda$6584      & $2.291 \times 10^{-7}$ \\
\midrule
O$^{+}$     & [O~\textsc{ii}] $\lambda$4085      & $1.120 \times 10^{-1}$ \\
            & [O~\textsc{ii}] $\lambda$4185      & $2.386 \times 10^{-2}$ \\
            & [O~\textsc{ii}] $\lambda$4346      & $1.072 \times 10^{-1}$ \\
O$^{2+}$    & [O~\textsc{iii}] $\lambda$4363      & $2.716 \times 10^{-4}$ \\
            & [O~\textsc{iii}] $\lambda$4959      & $9.314 \times 10^{-5}$ \\
            & [O~\textsc{iii}] $\lambda$5007      & $9.928 \times 10^{-5}$ \\
\midrule
Ne$^{3+}$   & [Ne~\textsc{iv}] $\lambda$3869      & $2.813 \times 10^{-5}$ \\
\midrule
Ar$^{2+}$   & [Ar~\textsc{iii}] $\lambda$7136     & $2.292 \times 10^{-7}$ \\
Ar$^{3+}$   & [Ar~\textsc{iv}] $\lambda$4711      & $4.878 \times 10^{-6}$ \\
            & [Ar~\textsc{iv}] $\lambda$4740      & $9.307 \times 10^{-7}$ \\
            & [Ar~\textsc{iv}] $\lambda$7237      & $9.210 \times 10^{-6}$ \\
            & [Ar~\textsc{iv}] $\lambda$7263      & $5.633 \times 10^{-6}$ \\
Ar$^{4+}$   & [Ar~\textsc{v}] $\lambda$7005       & $8.184 \times 10^{-7}$ \\
\midrule
S$^{+}$     & [S~\textsc{ii}] $\lambda$4069       & $3.972 \times 10^{-7}$ \\
            & [S~\textsc{ii}] $\lambda$6716       & $9.918 \times 10^{-8}$ \\
            & [S~\textsc{ii}] $\lambda$6731       & $7.954 \times 10^{-8}$ \\
\midrule
Cl$^{2+}$   & [Cl~\textsc{iii}] $\lambda$5518     & $1.424 \times 10^{-7}$ \\
            & [Cl~\textsc{iii}] $\lambda$5538     & $6.145 \times 10^{-8}$ \\
\bottomrule
\end{tabular}
\end{threeparttable}
\end{table}

\subsubsection{Total Elemental Abundances:}
In NGC 4361, we calculated the overall elemental abundances of He, O, N, Ne, Ar, S, Si, and Cl. This was done using the direct sum if specific ions are detected, and ICFs were used where only few ions of a species were detected. For argon, we include ions such as Ar$^{3+}$, Ar$^{4+}$, and Ar$^{5+}$. Since no Ar$^{2+}$ was detected, and we detected high-ionization stages, we used a direct sum method that fits this situation. The total abundance of helium was derived from the combination of He$^{+}$ and He$^{2+}$, both identified from the helium recombination lines. Only Ne$^{3+}$ was observed for neon, so we applied a correction based on the presence of oxygen ions, O$^{2+}$ and O$^{3+}$. The total oxygen abundance was calculated from the directly observed O$^{2+}$ and O$^{3+}$ ions. For Sulfur, we derived its abundance from only S$^{2+}$, using a correction factor that considered the observable oxygen and helium ions. Nitrogen calculation used ICF formula from \cite{TorresPeimbert1977}, suitable when only N$^{2+}$ is visible alongside oxygen ions. Chlorine abundance was estimated from just Cl$^{3+}$ ions, using a correction factor that considers the influence of helium ions, specifically He$^{2+}$, based on a method from \citep{Kwitter2001}. These calculations give us a clearer picture of the chemical composition of NGC 4361. The results, formatted as $12 + \log(\mathrm{X}/\mathrm{H})$, are detailed in a Table \ref{tab:elemental_abun_NGC4361} and compared with earlier report and the solar neighborhood.

\begin{table}[h!]
\centering
\begin{threeparttable}
\caption{Elemental Abundances in NGC 4361, presented in units of $12 + \log(\mathrm{X}/\mathrm{H})$\label{tab:elemental_abun_NGC4361}}
\small 
\begin{tabular}{lccc}
\toprule
\textbf{Element} & \textbf{This Work} & \textbf{Literature\textsuperscript{a}} & \textbf{Solar\textsuperscript{b}} \\
\midrule
He & 10.66 & 11.02 & 10.90 \\
N & 7.73 & 7.43 & 7.83 \\
O & 8.23 & 7.83 & 8.69 \\
Ne & 7.49 & 7.57 & 8.05 \\
S & 5.29 & --- & 7.19 \\
Cl & 5.49 & --- & 5.50 \\
Ar & 6.79 & 5.91 & 6.55 \\
\midrule
N/O & 0.32 & 0.40 & 0.14 \\
\bottomrule
\end{tabular}
\textsuperscript{a}\citet{Torres1990} \\
\textsuperscript{b}\citet{Lodders2003}
\end{threeparttable}
\end{table}

\section{Discussion}

\subsection{Physical Conditions and Ionization Structure}
NGC 2392 and NGC 4361 have significantly different physical conditions. In NGC 2392, the electron temperature starts around $T_e \sim 10^4$~K in regions of low ionization, based on [O~III] line diagnostics, and rises to approximately $1.8 \times 10^4$~K in zones of higher ionization, traced by [Ar~III]. The temperature continues to increase closer to the nebular core, reaching values above $2 \times 10^4$~K, where highly ionized species such as [O~IV] and [Ne~V] are found, this is one of the most extreme temperature gradients observed in any planetary nebula \citep{Zhang2012}. 

The inner shell of NGC 2392 also has relatively high densities, with electron densities around 3000~cm$^{-3}$ as derived from [S~II] and [S III] diagnostics. These values are consistent with an average density of roughly 2000~cm$^{-3}$ across the 18$''$ bright core \citep{Zhang2012, Pottasch2008}, indicating a compact and dense inner region.

On the other hand, NGC 4361 is characterized by hotter and more diffuse gas. Its typical electron temperature lies between $1.4 \times 10^4$ and $1.5 \times 10^4$~K, as derived from [O~III] lines \citep{Walsh2024}. This higher temperature reflects the central star's stronger and harder radiation field. The inner regions show electron densities ranging from 3500 to 4000~cm$^{-3}$ \citep{Walsh2024}, while earlier observations suggest that the more extended envelope is much less dense, around 1200--1500~cm$^{-3}$ \citep{Henry2000}. The nebula appears to have a two-layered structure: a dense inner region surrounded by a very diffuse halo, indicative of a more evolved planetary nebula.

These differences also reflect how each nebula is ionized. NGC 2392 has a medium to high excitation level, powered by a central star with an effective temperature of about 40,000~K \citep{Pauldrach2004}. The spectrum is dominated by [O~III] emission, while He~II lines are relatively weak, indicating that the nebula is mostly radiation-bounded. Its double-shell structure, a bright 18$''$ core and a fainter 40$''$ outer halo, supports a stratified ionization layout, where high-energy emission comes from the inner region, and lower-energy lines dominate the outer shell \citep{O'Dell2002}.

By contrast, NGC 4361 is an extremely high-excitation object. Its He~II $\lambda4686$ line is unusually strong, even brighter than H$\beta$ in many regions, placing it at the very top of the excitation class scale \citep{Aller1956, Dopita1990}. The nebula is essentially fully ionized, with very little neutral gas remaining. Faint traces of [N~II] and [O~II] emission were only recently discovered in over a hundred small, compact clumps scattered across the nebula, known as “Freckles” \citep{Walsh2024}. These knots are slightly cooler ($T_e \sim 11,000$~K) and have lower ionization than the surrounding gas, but their overall contribution to the nebula’s emission is small. Thus, NGC 4361 is fundamentally different from NGC 2392; it is more evolved nebula dominated by high-energy radiation.

\subsection{Elemental Abundances and Chemical Properties}
The chemical compositions of NGC~2392 and NGC~4361 offer clear insights into the histories of their progenitor stars. 


Although both NGC 2392 and NGC 4361 have N/O ratios consistent with Type II classification, their measured elemental abundances, specifically subsolar values of O/H and Ne/H, suggest progenitors with metallicities below solar. This indicates formation epochs or locations within the Galaxy where interstellar oxygen was not yet enriched to solar levels. Therefore, for the progenitor of NGC 2392, a possible scenario is that it formed after the Sun, but in a low-metallicity region where the oxygen abundance was much lower than the solar value. This interpretation reconciles its evolutionary timescale with the observed nebular properties and provides a natural explanation for its subsolar O/H. Similarly, the extremely low metallicity and N/O ratio of NGC 4361 further confirm its Population II origin, placing its progenitor in an older, low-mass stellar population.

Carbon abundance in NGC~2392 is more uncertain. Optical C~II lines are weak in our spectra, but mid-infrared data suggest a C/O ratio near or slightly above unity \citep{Pottasch2008}, indicating that some carbon enrichment may have occurred. The neon, sulfur, and argon abundances track oxygen closely, suggesting a uniform chemical composition consistent with a moderately metal-poor origin, possibly in the outer Galactic disk \citep{Pottasch2008}.
Upon inspection, the IRS spectrum of NGC 2392 does not show a prominent SiC emission feature near $11.3\mu$m or clear SiO signatures, which have also been noted in previous studies. The absence or weakness of these dust features suggests that NGC 2392 may not be strongly carbon-rich $(C/O<1)$, even if nebular emission line diagnostics indicate moderate carbon enrichment. This aligns with our finding of 
$12 + log(C/H) = 8.31$ and 
$12 + log(O/H) = 8.75$, yielding $C/O = 0.36$(by number). The chemical dust composition supports the spectroscopic result that the nebula is either marginally carbon-rich or near the C/O threshold. Our interpretation is consistent with mid-infrared studies that find no clear PAH or SiC features in NGC 2392, further suggesting a moderate C/O ratio.

In contrast, NGC~4361 exhibits a more chemically depleted composition, characteristic of an older stellar population. Its oxygen abundance is similar to that of NGC~2392 ($12 + \log(\text{O/H}) \approx 8.3$), but here it reflects a metal-poor initial composition rather than processed material. NGC~4361 has long been classified as a Population~II PN \citep{Torres1990}. Its nitrogen abundance is extremely low (N/O $\approx$ 0.03--0.08), indicating minimal nitrogen processing, pointing to a low-mass, metal-poor progenitor born with intrinsically low N and O \citep{Henry2000}.

The helium abundance in NGC~4361 is similar to that of NGC~2392 (He/H~$\approx$~0.10), just slightly above primordial, suggesting minimal helium enrichment during the asymptotic giant branch (AGB) phase. Together with its high Galactic latitude and kinematic properties, NGC~4361 is placed among halo or thick-disk objects formed in a low-metallicity environment \citep{Henry2000}.

Heavy elements in NGC~4361 are also underabundant. Sulfur, for example, is approximately 25 times lower than in NGC~2392, though exact values depend on uncertain ionization correction factors due to the nebula’s high excitation \citep{Henry2000}. Neon and argon are also subsolar and scale with oxygen, as expected in a metal-poor PN.

NGC~4361 hosts an O(H)6 type central star \citep{GonzalezSantamaria2021}, usually associated with strong helium and carbon enrichment, but the nebular carbon abundance shows only mild enhancement. Deep optical spectroscopy indicates a fairly uniform C/O ratio across the nebula, elevated only by about 0.2~dex compared to its metallicity \citep{Liu1998}. This suggests that the central star experienced only a few thermal pulses, and any associated carbon or s-process enrichment was moderate.

\subsection{Morphological Differences and Ionization Structure}
The formation of planetary nebulae (PNe) is governed by the interaction between the fast, tenuous stellar wind from the evolving central star and the slower, dense 'superwind' ejected during the terminal asymptotic giant branch (AGB) phase. This interaction generates a high-temperature (hot) bubble at the center, which cools inefficiently and expands outward, sweeping up and compressing the previously expelled AGB wind. The resultant structures, the inner hot bubble and the swept-up shell, shape both the dynamical evolution and observed morphology of the nebulae \citep{Perinotto2004}. This 'interacting stellar winds' scenario is well-supported for both NGC 2392 and NGC 4361 and provides the physical foundation for interpreting their observed characteristics.

The contrasting physical conditions of NGC~2392 and NGC~4361 are clearly reflected in their morphologies and ionization structures. The layered structure of NGC 2392 is believed to result from two episodes of mass loss, a slow, dense wind during the AGB phase forming the outer shell, later swept up by a faster post-AGB wind to create the inner bubble. High-resolution imaging reveals that the inner shell is filled with radial filaments and clumpy features resembling comet tails, giving rise to the “Eskimo” appearance \citep{O'Dell2002}. Despite this complexity, the nebula maintains an overall spherical symmetry. Kinematic studies confirm that the inner shell is expanding faster (about 70~km~s$^{-1}$) than the outer halo (around 15~km~s$^{-1}$), consistent with a wind-driven bubble model \citep{O'Dell2002}. This morphology also aligns well with the nebula's ionization structure. High-ionization lines like [O~III], [Ne~III], and He~II are strongest in the inner shell, while the outer halo emits mainly low-ionization lines such as [N~II] and [O~II]. This spatial stratification results in a fully ionized inner region surrounded by a recombination zone marking the nebula’s ionization boundary. For example, the [N~II]~$\lambda$6584/H$\alpha$ ratio increases outward, indicating a transition to N$^+$ in the outer layers. The presence of moderate-density knots in the halo, visible in optical images, suggests that some neutral or dusty material remains, being slowly photoionized at the edges \citep{O'Dell2002}. All these features point to NGC~2392 being a relatively young, radiation-bounded nebula that is still evolving.

NGC~4361 displays a more diffuse, elliptical shape with faint asymmetric extensions. Ground-based and deep imaging reveal curved “arms” stretching to the northeast and southwest, giving the nebula a four-lobed or cloverleaf appearance, earning it the nickname “Lawn Sprinkler Nebula.” Unlike NGC~2392, there is no clear boundary between an inner and outer shell; instead, its surface brightness and ionization decline gradually with radius. This suggests that the nebula has entered a more evolved stage, where initial shell structures have merged or dispersed.

The nebula remains highly ionized even out to its outermost regions, indicating that it is matter-bounded, with very little neutral material remaining to absorb the ionizing photons. A striking morphological feature of NGC~4361 is the discovery of over a hundred compact, low-ionization knots scattered across its face \citep{Walsh2024}. These structures resemble the cometary knots seen in other evolved PNe, like the Helix \citep{Meaburn1998}, but in NGC~4361, they are more symmetric and often arranged in linear chains. This configuration hints that NGC~4361's progenitor may have ejected material preferentially in the equatorial plane, either due to stellar rotation or due to the influence of a binary companion, which has fragmented into the observed knots. The bulk of the nebula consists of highly ionized, low-density gas filling void volume. He~II emission remains strong even at large distances from the center, confirming the nebula's fully ionized state. Altogether, NGC~4361 represents an older, more evolved planetary nebula in the final stages of its visible lifetime, tenuous, extended, and largely transparent to ionizing radiation.

\subsection{Zanstra temperature and Evolution of NGC 2392 and NGC 4361 on the H–R diagram}

The Helium Zanstra temperature ($T_{\text{Zanstra, He}}$) provides a reliable estimate of the effective temperature of central stars, particularly suitable for high-excitation planetary nebulae such as NGC~2392 and NGC~4361. We calculated the Helium Zanstra temperatures using the well-known Zanstra method, which is based on the observed flux ratio of the He\,\textsc{ii} $\lambda4686~\text{\AA}$ recombination line relative to H$\beta$ ($\lambda4861~\text{\AA}$). The Zanstra method assumes a blackbody radiation field for the central star and ionization-bounded conditions for the nebula.

Emission line intensities (relative to H$\beta = 100$, intrinsic fluxes; from Table \ref{tab:A1} and \ref{tab:A2}):
\begin{itemize}
  \item NGC~2392: He\,\textsc{ii}~$\lambda4686 = 46.0$, H$\beta = 100.0$
  \item NGC~4361: He\,\textsc{ii}~$\lambda4686 = 110.0$, H$\beta = 100.0$
\end{itemize}
\textbf{He\,\textsc{ii}~$\lambda4686$/H$\beta$ Ratios:}
\begin{itemize}
\item NGC~2392: 0.46
  \item NGC~4361: 1.10
\end{itemize}

Using standard calibrations of the He\,\textsc{ii}~$\lambda4686$/H$\beta$ intensity ratio versus Zanstra temperature (see, \citep{Osterbrock2006}), the temperatures are estimated as follows:
\begin{itemize}
  \item NGC~2392: $\approx 88,000$~K
  \item NGC~4361: $\approx 114,000$~K
\end{itemize}
In, NGC~2392 He\,\textsc{ii} Zanstra temperature is approximately 88,000~K (73,000~K previously estimated by \citep{Pottasch2008}), indicating a hot central star capable of producing significant He\,\textsc{ii} emission, though it remains below the temperatures typically associated with the most extreme [WC] or O(H)-type central stars of planetary nebulae (CSPNe).
Whereas in NGC~4361, the central star exhibits a considerably higher Zanstra temperature, approximately 114,000~K, consistent with the presence of a dense He$^{++}$ region and strong overall nebular excitation. This aligns well with the extreme ionization characteristics previously noted for this nebula.

\begin{figure}
    \centering
    \includegraphics[width=1\linewidth]{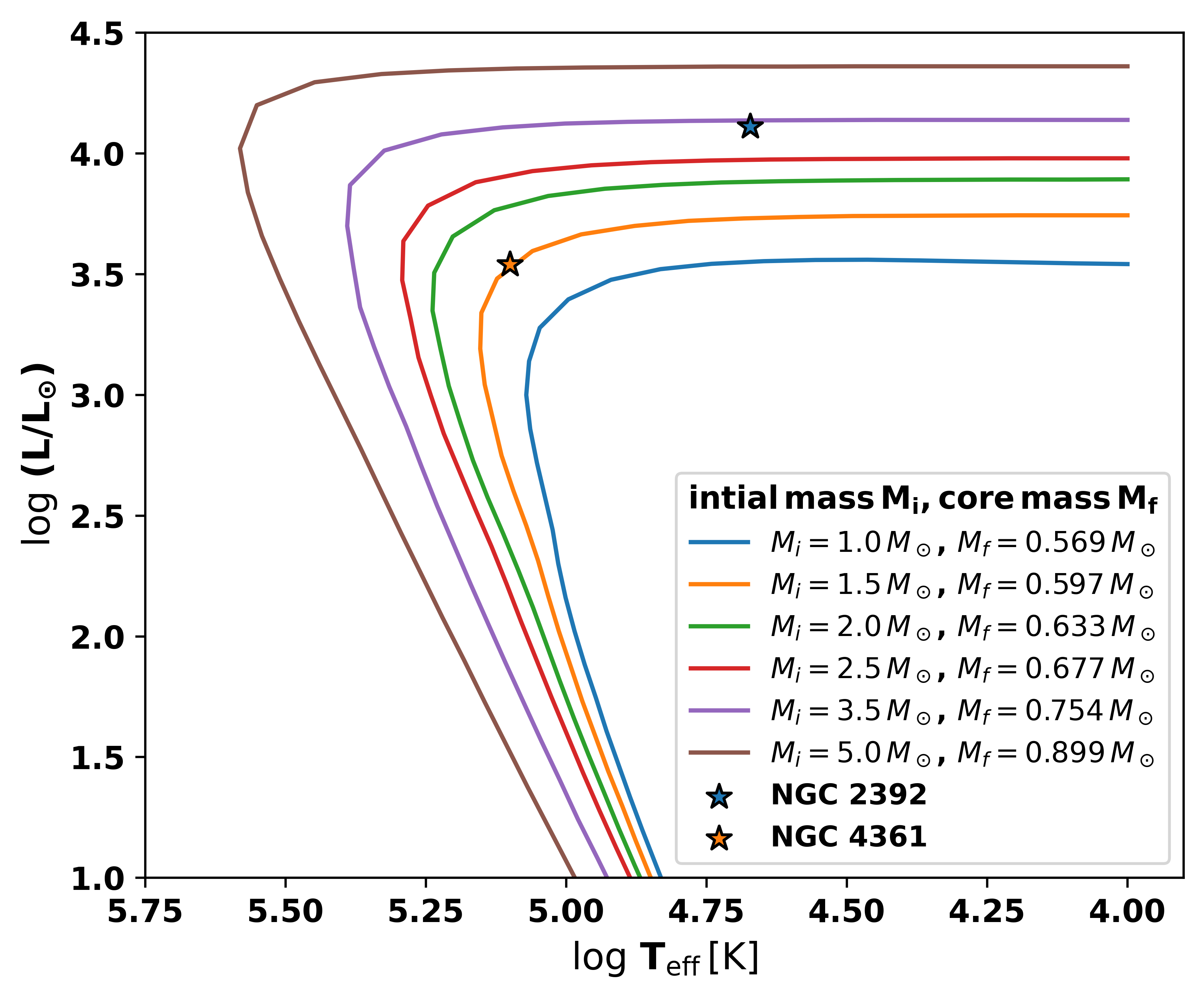}
    \caption{H–R diagram showing post-AGB evolutionary tracks from \citet{VassiliadisWood1994} 
for $Z=0.016$ and various core masses. We place NGC 2392 and
NGC 4361 among the model tracks at the coordinates $\log(L/L_\odot)$,
Log $T_{\mathrm{eff}}$ corresponding to our estimated values of $T_{\mathrm{eff}}$ and L.}

    \label{fig:HR_tracks}
\end{figure}

Figure~\ref{fig:HR_tracks} overlays both objects on the \citet{VassiliadisWood1994} post-AGB H-burning tracks computed for solar-neighborhood metallicity ($Z=0.016$).\footnote[7]{The coloured curves are the public grids kindly provided by P. Vassiliadis.}
Central star effective temperatures are derived from non-LTE stellar atmosphere analyses: 
$T_{\mathrm{eff}} = 47$\,kK for NGC\,2392 \citep{Pauldrach2004} and $T_{\mathrm{eff}} = 126$\,kK for NGC\,4361 \citep{GonzalezSantamaria2021}. The factor-of-two temperature difference immediately suggests these objects occupy distinct evolutionary phases, with NGC\,4361's higher temperature indicative of a more evolved state.
We adopt \textit{Gaia}\,EDR3 distances of $d = 1.73\pm0.15$\,kpc for NGC\,2392 and $d = 1.01\pm0.07$\,kpc for NGC\,4361 \citep{GonzalezSantamaria2021}. Extinction corrections are minimal for both objects, consistent with their moderate Galactic latitudes. For NGC\,4361, we adopt $E(B\!-\!V)=0.05\pm0.01$ and $A_{V}=0.07\pm0.02$\,mag \citep{GonzalezSantamaria2021}, while NGC\,2392 exhibits similarly low reddening typical of nearby planetary nebulae \citep{GonzalezSantamaria2021}.
Central-star $V$ magnitudes are $10.50\pm0.05$ for NGC 2392 \citep{Pauldrach2004} and $13.21\pm0.05$ for NGC\,4361 (DeepSkyCorner CSPN catalogue). Bolometric corrections appropriate to the adopted effective temperatures, $BC_V=-4.5\pm0.2$ for the 47\,kK star \citep{Flower1996} and $BC_V=-7.5\pm0.3$ for the 126\,kK object \citep{Rauch2003}.
Following standard distance-luminosity relations:
\begin{equation}
    M_V = m_V - 5\log_{10}\!\left(\frac{d}{10\,\mathrm{pc}}\right) - 3.1E(B\!-\!V)
\end{equation}
\begin{equation}
    M_{\rm bol} = M_V + BC_V
\end{equation}
and from $M_{\rm bol,\odot}=4.74$\,mag, we obtain stellar luminosities via
\begin{equation}
    \log\!\left(\frac{L}{L_\odot}\right) = -0.4\,(M_{\rm bol}-M_{\rm bol,\odot})
\end{equation}
yielding $\log(L/L_\odot) = 4.11$ for NGC 2392 and $\log(L/L_\odot) = 3.54$ for NGC 4361.

The H–R diagram placement reveals fundamentally different evolutionary states for these two objects. NGC 2392 sits just below on the horizontal plateau of the $0.75\,M_\odot$ core mass post-AGB track. The stellar core is still in its high-luminosity, constant-radius heating phase, undergoing H-shell burning at nearly constant luminosity while the effective temperature gradually increases \citep{VassiliadisWood1994}. The star has not yet reached the maximum temperature point that precedes the turn-down toward the white dwarf cooling sequence. This evolutionary phase typically lasts $\sim$10$^4$\,years \citep{MillerBertolami2016}, during which the ionizing photon flux steadily hardens.

By contrast, NGC\,4361 lies on the early cooling branch of the $0.60\,M_\odot$ core mass track, passing through the luminosity peak and begun its irreversible descent toward the white dwarf regime. This placement is consistent with its significantly higher effective temperature (126\,kK vs. 47\,kK) and the nebula's extreme ionization properties, including He\,{\sc ii}/H$\beta$ line ratios exceeding 10, the maximum value in the \citet{Dopita1990} classification scheme, and the presence of highly ionized species such as [Ne\,{\sc v}]\,3426\,\AA\ \citep{Aller1978}.

The H–R diagram therefore corroborates independent spectroscopic evidence that NGC\,4361 is the more evolved, optically thin planetary nebula, whereas NGC\,2392 is several thousand years ``younger'' in post-AGB terms and still photoionization-bounded.
NGC\,2392 remains ionization-bounded (optically thick), with sufficient circumstellar material to absorb essentially all Lyman continuum photons. The nebula exhibits typical stratified ionization structure with detectable low-ionization features including [N\,{\sc ii}] emission and clear boundaries between ionized and neutral regions. 
NGC\,4361, conversely, represents a matter-bounded (optically thin) system where the ionizing photon rate exceeds the local recombination rate due to the low density of the highly dispersed envelope. The nebula's extreme He\,{\sc ii}\,4686\,\AA\ strength, weak He\,{\sc i} emission, and very high electron temperatures ($T_e \sim 18,000$\,K; \citet{Torres1990, Liu1998}) all point to a rarefied, matter-bounded configuration.

Recent MUSE integral field observations confirm this interpretation through the detection of compact low-ionization ``freckles'' embedded within predominantly high-ionization gas \citep{Walsh2024}, suggesting a complex, inhomogeneous density structure characteristic of evolved planetary nebulae where the stellar wind has dispersed most of the circumstellar envelope.

\subsection{PAH Emission in NGC~2392 and NGC~4361}
Polycyclic aromatic hydrocarbons (PAHs) produce characteristic infrared (IR) emission features, such as the 3.3\,$\mu$m C--H stretching mode, observed in a variety of astrophysical environments \citep{Tielens2008, Rastogi2013}. However, in our narrow-band 3.28\,$\mu$m observations of NGC~2392 and NGC~4361 using the 3.6-m DOT, we found no detectable PAH signatures. The absence of PAH emission in both nebulae suggests either the complete absence or a very low abundance of these molecules, below the sensitivity of our observations. Indeed, previous studies of young, carbon-rich PNe such as NGC~7027 and BD+30$^{\circ}$3639 using the same TIRCAM2 instrument on the DOT telescope revealed distinct PAH emission localized to cooler, neutral regions outside the ionized core \citep{Anand2020}. These observations demonstrate that PAHs survive preferentially in environments shielded from intense stellar radiation. 


For NGC 2392, the absence of PAH features may be attributed not only to the lack of carbon-rich chemistry but also to the nebula's radiation-bounded nature. PAHs can survive only in regions where the local photon energies are below about 20 eV, which restricts their presence to peripheral zones of the ionized gas or in PDRs. In density-bounded nebulae, molecules may persist in shielded clumps, while in ionization-bounded nebulae, the advancing ionization front confines the gas and allows PAHs to form at the interface with the molecular envelope, provided the nebula is carbon-rich. In the case of NGC 2392, although its C/O ratio is close to unity, the inefficient formation of PAHs and the lack of a strongly C-rich chemistry explain the absence of observable PAH features.

For NGC~4361, this can be explained by its highly ionized environments, which tend to inhibit the survival of PAH molecules. PAHs are susceptible to destruction via ionization and photodissociation processes in regions with intense ultraviolet radiation \citep{Tielens2008}. Additionally, the formation of PAHs strongly depends on carbon-rich conditions (high C/O ratios), typically resulting from third dredge-up events during the late AGB phase \citep{Tielens2008}. Hence, nebulae originating from less carbon-rich progenitors, or nebulae lacking protected neutral regions, are not expected to exhibit prominent PAH emission. 

Laboratory and modeling studies further support this scenario. Vibrational spectroscopy of PAHs shows that their IR emission profiles vary significantly with molecular structure and ionization state \citep{Maurya2012}. Modeling the observed interstellar aromatic infrared bands (AIBs) indicates that these emissions result from complex mixtures of PAH molecules of various sizes and ionization levels \citep{Pathak2008}. Under intense UV radiation, such as that present in highly ionized nebulae, PAHs become predominantly ionized or fragmented, leading to weak or absent IR emission features at 3.3\,$\mu$m.

Although our ground-based observations failed to detect PAH emission in NGC~2392 and NGC~4361, recent JWST mid-infrared observations highlight the potential of advanced instruments to detect elusive PAH signatures in PNe. JWST/MIRI spectra have successfully identified clear PAH bands in carbon-rich PNe like SMP~LMC~058 \citep{Jones2023}, demonstrating JWST's superior sensitivity and spectral resolution compared to previous facilities like \textit{Spitzer}.

Future observations with JWST may thus offer deeper insights into the presence or absence of PAHs in objects like NGC~2392 and NGC~4361. At present, our non-detection aligns well with theoretical expectations and prior observational constraints, underscoring how local physical and chemical conditions critically govern the presence and detectability of PAHs in PNe.

\section{Conclusion}

NGC~2392 and NGC~4361 offer a powerful comparative case study of PNe, illustrating how differences in progenitor mass, age, and evolutionary history manifest in their physical and chemical properties. NGC~2392, with its moderate excitation, distinct double-shell structure, and relatively young central star, represents a classic example of a radiation-bounded nebula in an early post-AGB phase. In contrast, NGC~4361 exhibits extremely high excitation and a matter-bounded structure, consistent with an older, metal-poor Population~II progenitor whose central star has evolved into a very hot O(H)6 type nucleus.

Physically, NGC~2392’s inner regions show typical electron temperatures around $T_e \sim 10^4$~K and densities of $N_e \sim 10^3$~cm$^{-3}$. NGC~4361 reaches higher temperatures ($T_e \approx 1.5 \times 10^4$~K) and maintains core densities of several~$\times 10^3$~cm$^{-3}$, consistent with its more advanced evolutionary stage where its central star’s radiation has ionized nearly all surrounding gas.

Chemically, both nebulae reflect their low-mass stellar origins, showing no high helium or nitrogen enrichment characteristic of massive progenitors. However, NGC~2392 shows moderate enrichment in helium, carbon, and nitrogen typical of a low-mass, thin-disk star with a slightly higher metallicity (Galactic anticenter region). NGC~4361, on the other hand, stands out for its significantly lower oxygen and nitrogen abundances, confirming its connection to an older, metal-poor Population~II progenitor. It returns moderate amounts of helium and carbon, and possibly small enrichments of neutron-capture elements like fluorine, to the ISM.

Together, these two nebulae demonstrate the diverse evolutionary paths of low- and intermediate-mass stars. NGC~2392 contributes modest enrichment to the interstellar medium at roughly half-solar metallicity, while NGC~4361 reflects chemical yields more representative of early Galactic environments.

The comparison between NGC~2392 and NGC~4361 highlights the close connection between stellar evolution and the physical state of PNe. Electron temperature, density, and ionization structure reflect the central star’s properties and history, while the nebular abundances preserve a record of the star’s initial composition and internal nucleosynthesis. These two objects may be viewed as complementary stages, perhaps even ``before and after'' snapshots, in PN evolution. Continued observations and modeling will further illuminate their nature, offering the opportunity to watch these cosmic structures transform and fade. In doing so, we gain deeper insights into the fate of Sun-like stars and the processes that enrich and shape the Galaxy over time.

\section*{Acknowledgements}
The authors gratefully acknowledge the reviewers for their constructive feedback and valuable suggestions, which helped improve the quality of this work.

We acknowledge the Himalayan Chandra Telescope (HCT) operated by the Indian Institute of Astrophysics (IIA) for providing the telescope time necessary for our observations. We also extend our thanks to the 3.6\,m Devasthal Optical Telescope (DOT) team operated by the Aryabhatta Research Institute of Observational Sciences (ARIES) for facilitating the TIRCAM2 observations. A.K.S acknowledges the financial support provided by the National Fellowship for Other Backward Classes (NFOBC), which the Ministry of Social Justice and Empowerment, Government of India, awarded. A.G. acknowledges support from DGAPA-UNAM and PAPIIT project IN105225.\\
This work has greatly benefited from the use of archival data obtained from the Mikulski Archive for Space Telescopes (MAST). MAST is operated by the Space Telescope Science Institute (STScI), and we utilized observations from the Hubble Space Telescope (HST) and the Pan-STARRS1 (PS1) Surveys hosted therein.\\


\onecolumn
\appendix
\renewcommand{\thetable}{A\arabic{table}}
\setcounter{table}{0}

\section{Optical Region Line Intensities}

\setlength{\LTcapwidth}{\textwidth} 
\scriptsize
\renewcommand{\arraystretch}{2}
\begin{longtable}{cc@{\extracolsep{0.1cm}}ccccccccc}

\caption{Optical region line intensities in NGC 2392 on a scale where H$\beta = 100$} \label{tab:opt_lines_NGC2392} 
\\
\toprule
$\lambda_{obs}$ & $\lambda_{rest}$ & $F \left( \lambda \right)$ & $I \left( \lambda \right)$ & Ion & Multiplet & Lower term & Upper term & g$_1$ & g$_2$ \\
\midrule
\endfirsthead
\toprule
$\lambda_{obs}$ & $\lambda_{rest}$ & $F \left( \lambda \right)$ & $I \left( \lambda \right)$ & Ion & Multiplet & Lower term & Upper term & g$_1$ & g$_2$ \\
\midrule
\endhead
\midrule
\multicolumn{10}{r}{\emph{Continued on next page}} \\
\bottomrule
\endfoot
\bottomrule
\endlastfoot
  3554.43 &   3554.42 &    5.399 $\pm$   0.739&   5.528 $^{  +0.853}_{  -0.883}$ & He I       &V34         &2p 3P*      &10d 3D               & 9         & 15 \\
  3568.51 &   3568.50 &    2.772 $\pm$   0.846&   3.116 $^{  +0.958}_{  -0.998}$ & Ne II      &V9          &3s' 2D      &3p' 2F*              & 6         &  8 \\
  3587.29 &   3587.28 &    1.300 $\pm$   0.501&   2.528 $^{  +0.565}_{  -0.580}$ & He I       &V31         &2p 3P*      &9d 3D                & 9         & 15 \\
  3634.26 &   3634.25 &    4.475 $\pm$   0.331&   4.565 $^{  +0.408}_{  -0.448}$ & He I       &V28         &2p 3P*      &8d 3D                & 9         & 15 \\
  3721.64 &   3721.63 &   67.806 $\pm$   1.880&  74.800 $^{  +3.600}_{  -3.800}$ & [S III]    &F2          &3p2 3P      &3p2 1S               & 3         &  1 \\
  3726.04 &   3726.03 &   54.976 $\pm$   1.874&  61.600 $^{  +3.200}_{  -3.400}$ & [O II]     &F1          &2p3 4S*     &2p3 2D*              & 4         &  4 \\
  3728.83 &   3728.82 &   10.258 $\pm$   1.869&  13.500 $^{  +2.100}_{  -2.100}$ & [O II]     & F1         & 2p3 4S*    & 2p3 2D*             & 4         &  6 \\
  3835.40 &   3835.39 &    1.663 $\pm$   0.566&   2.136 $^{  +0.626}_{  -0.621}$ & H I        &H9          &2p+ 2P*     &9d+ 2D               & 8         &162 \\
  3862.61 &   3862.60 &   11.014 $\pm$   4.889&  17.500 $^{  +5.300}_{  -5.400}$ & Si II      &V1          &3p2 2D      &4p 2P*               & 4         &  2 \\
  3868.76 &   3868.75 &  110.594 $\pm$   4.824& 120.000 $^{  +7.000}_{  -7.000}$ & [Ne III]   &  F1        & 2p4 3P     & 2p4 1D              & 5         &  5 \\
  3888.66 &   3888.65 &   19.092 $\pm$   1.404&  20.300 $^{  +1.600}_{  -1.800}$ & He I       &V2          &2s 3S       &3p 3P*               & 3         &  9 \\
  3907.47 &   3907.46 &    5.598 $\pm$   1.216&   6.230 $^{  +1.330}_{  -1.360}$ & O II       &V11         &3p 4D*      &3d 4P                & 6         &  6 \\
  3967.47 &   3967.46 &   42.724 $\pm$   1.713&  46.900 $^{  +2.400}_{  -2.500}$ & [Ne III]   &F1          &2p4 3P      &2p4 1D               & 3         &  5 \\
  3970.08 &   3970.07 &   40.051 $\pm$   1.618&  44.700 $^{  +2.300}_{  -2.400}$ & H I        &H7          &2p+ 2P*     &7d+ 2D               & 8         & 98 \\
  4060.61 &   4060.60 &    2.676 $\pm$   0.616&   2.877 $^{  +0.668}_{  -0.669}$ & O II       &V97         &3d 2F       &4f 2[4]*             & 8         &  0 \\
  4068.61 &   4068.60 &    2.326 $\pm$   0.790&   3.035 $^{  +0.851}_{  -0.870}$ & [S II]     &F1          &2p3 4S*     &2p3 2P*              & 4         &  4 \\
  4076.36 &   4076.35 &    2.754 $\pm$   0.662&   2.735 $^{  +0.704}_{  -0.727}$ & [S II]     &F1          &2p3 4S*     &2p3 2P*              & 2         &  4 \\
  4101.75 &   4101.74 &   21.508 $\pm$   2.760&  26.400 $^{  +2.900}_{  -2.900}$ & H I        & H6         & 2p+ 2P*    & 6d+ 2D     	        & 8         & 72 \\
  4110.79 &   4110.78 &    5.288 $\pm$   2.596&   9.200 $^{  +2.790}_{  -2.810}$ & O II       &V20         &3p 4P*      &3d 4D                & 4         &  2 \\
  4341.24 &   4340.47 &   41.794 $\pm$   1.186&  43.600 $^{  +1.300}_{  -1.400}$ & H I        &  H5        & 2p+ 2P*    & 5d+ 2D              & 8         & 50 \\
  4346.32 &   4345.55 &    9.641 $\pm$   1.211&  10.200 $^{  +1.300}_{  -1.300}$ & O II       &  V2        & 3s 4P      & 3p 4P*              & 4         &  2 \\ 
  4363.98 &   4363.21 &   11.783 $\pm$   0.829&  13.528 $^{  +0.882}_{  -0.943}$ & [O III]    &F2          &2p2 1D      &2p2 1S               & 5         &  1 \\
  4367.66 &   4366.89 &    8.716 $\pm$   0.724&   8.539 $^{  +0.746}_{  -0.817}$ & N III      &V2          &3s 4P       &3p 4P*               & 6         &  4 \\
  4379.88 &   4379.11 &    2.537 $\pm$   0.530&   1.672 $^{  +0.559}_{  -0.550}$ & N III      &V18         &4f 2F*      &5g 2G                &14         & 18 \\
  4388.70 &   4387.93 &    2.795 $\pm$   0.513&   2.180 $^{  +0.521}_{  -0.536}$ & He I       &V51         &2p 1P*      &5d 1D                & 3         &  5 \\
  4658.92 &   4658.10 &    4.528 $\pm$   0.344&   4.267 $^{  +0.341}_{  -0.371}$ & [Fe III]   &F3          &3d6 5D      &3d6 3F2              & 9         &  9 \\
  4686.51 &   4685.68 &   47.282 $\pm$   1.521&  46.000 $^{  +1.600}_{  -1.600}$ & He II      &  3.4       & 3d+ 2D     & 4f+ 2F*             &18         & 32 \\
  4712.20 &   4711.37 &    4.040 $\pm$   0.596&   4.730 $^{  +0.578}_{  -0.659}$ & [Ar IV]    &F1          &3p3 4S*     &3p3 2D*              & 4         &  6 \\
  4740.45 &   4740.17 &    3.233 $\pm$   0.903&   3.574 $^{  +0.941}_{  -0.916}$ & [Ar IV]    &F1          &3p3 4S*     &3p3 2D*              & 4         &  4 \\
  4754.97 &   4754.69 &    1.916 $\pm$   0.492&   1.549 $^{  +0.500}_{  -0.498}$ & [Fe III]   &F 3         &3d6 5D      &3d6 3F               & 7         &  9 \\
  4788.41 &   4788.13 &    1.927 $\pm$   0.277&   1.626 $^{  +0.277}_{  -0.276}$ & N II       &V20         &3p 3D       &3d 3D*               & 5         &  5 \\
  4861.61 &   4861.33 &  105.065 $\pm$   5.804& 100.000 $^{  +6.000}_{  -6.000}$ & H I        &  H4        & 2p+ 2P*    & 4d+ 2D              & 8         & 32 \\
  4959.20 &   4958.91 &  331.336 $\pm$  22.129& 346.00 $^{ +22.000}_{ -22.000}$  & [O III]    &  F1        & 2p2 3P     & 2p2 1D              & 3         &  5 \\
  5007.13 &   5006.84 & 1083.361 $\pm$  51.477&1040.00  $^{ +50.000}_{ -50.000}$ & [O III]    &  F1        & 2p2 3P     & 2p2 1D              & 5         &  5 \\
  5048.04 &   5047.74 &    5.260 $\pm$   0.616&   4.361 $^{  +0.610}_{  -0.588}$ & He I       &V47         &2p 1P*      &4s 1S                & 3         &  1 \\
  5192.12 &   5191.82 &    0.903 $\pm$   0.122&   0.889 $^{  +0.119}_{  -0.118}$ & [Ar III]   &F3          &2p4 1D      &2p4 1S               & 5         &  1 \\
  5270.71 &   5270.40 &    2.411 $\pm$   0.176&   2.161 $^{  +0.173}_{  -0.173}$ & [Fe III]   &  F1        & 3d6 5D     & 3d6 3P2             & 7         &  5 \\
  5410.86 &   5411.52 &    1.917 $\pm$   0.172&   1.713 $^{  +0.159}_{  -0.176}$ & He II      &4.7         &4f+ 2F*     &7g+ 2G               &32         & 98 \\
  5516.99 &   5517.66 &    1.205 $\pm$   0.300&   1.540 $^{  +0.283}_{  -0.286}$ & [Cl III]   &F1          &2p3 4S*     &2p3 2D*              & 4         &  6 \\
  5536.93 &   5537.60 &    2.969 $\pm$   0.346&   2.362 $^{  +0.335}_{  -0.327}$ & [Cl III]   &F1          &2p3 4S*     &2p3 2D*              & 4         &  4 \\
  5753.90 &   5754.60 &    1.414 $\pm$   0.139&   1.244 $^{  +0.129}_{  -0.144}$ & [N II]     &F3          &2p2 1D      &2p2 1S               & 5         &  1 \\
  5874.95 &   5875.66 &   12.419 $\pm$   0.373&  10.954 $^{  +0.511}_{  -0.511}$ & He I       &  V11       & 2p 3P*     & 3d 3D               & 9         & 15 \\
  6298.66 & 6300.34 & 2.59 $\pm$ 0.050 & 2.58 $^{ +0.045}_{ -0.055}$ & [O I] & F1 & 2p4 3P & 2p4 1D & 5 & 5 \\
  6311.96 &   6312.10 &    2.996 $\pm$   0.372&   2.984 $^{  +0.343}_{  -0.387}$ & [S III]    &F3          &2p2 1D      &2p2 1S               & 5         &  1 \\
  6547.94 &   6548.10 &   27.335 $\pm$   4.918&  23.200 $^{  +4.400}_{  -4.500}$ & [N II]     &F1          &2p2 3P      &2p2 1D               & 3         &  5 \\
  6562.61 &   6562.77 &  265.687 $\pm$  18.549& 262.000 $^{  +9.000}_{  -9.000}$ & H I        &  H3        & 2p+ 2P*    & 3d+ 2D              & 8         & 18 \\
  6583.34 &   6583.50 &   89.088 $\pm$   6.274&  79.000 $^{  +6.800}_{  -6.800}$ & [N II]     &  F1        & 2p2 3P     & 2p2 1D              & 5         &  5 \\
  6678.00 &   6678.16 &    2.314 $\pm$   0.274&   2.301 $^{  +0.255}_{  -0.286}$ & He I       &V46         &2p 1P*      &3d 1D                & 3         &  5 \\
  6716.28 &   6716.44 &    6.120 $\pm$   0.437&   5.162 $^{  +0.457}_{  -0.501}$ & [S II]     &F2          &2p3 4S*     &2p3 2D*              & 4         &  6 \\
  6730.66 &   6730.82 &    9.301 $\pm$   0.682&   7.605 $^{  +0.702}_{  -0.773}$ & [S II]     &F2          &2p3 4S*     &2p3 2D*              & 4         &  4 \\
  7065.25 &   7065.25 &    4.154 $\pm$   0.304&   3.282 $^{  +0.320}_{  -0.354}$ & He I       &V10         &2p 3P*      &3s 3S                & 9         &  3 \\
  7135.80 &   7135.80 &   17.173 $\pm$   1.509&  15.500 $^{  +1.600}_{  -1.700}$ & [Ar III]   &F1          &3p4 3P      &3p4 1D               & 5         &  5 \\
  7751.26 &   7751.06 &    3.987 $\pm$   0.395&   3.700 $^{  +0.413}_{  -0.465}$ & [Ar III]   &            &3p4 3P      &3p4 1D               & 3         &  5 \\
  9067.80 &   9068.60 &   29.210 $\pm$   4.244&  20.700 $^{  +3.900}_{  -3.900}$ & [S III]    &            & 3p2 3P     & 3p2 1D              & 3         &  5 \\
\label{tab:A1}
\end{longtable}

\begin{longtable}{cc@{\extracolsep{0.1in}}ccccccccc}
\caption{Optical region line intensities in NGC 4361 on a scale where H$\beta = 100$} 
\label{tab:linelist_NGC4361} \\
\toprule
$\lambda_{obs}$ & $\lambda_{rest}$ & $F \left( \lambda \right)$ & $I \left( \lambda \right)$ & Ion & Multiplet & Lower term & Upper term & g$_1$ & g$_2$ \\
\midrule
\endfirsthead
\toprule
$\lambda_{obs}$ & $\lambda_{rest}$ & $F \left( \lambda \right)$ & $I \left( \lambda \right)$ & Ion & Multiplet & Lower term & Upper term & g$_1$ & g$_2$ \\
\midrule
\endhead
\midrule
\multicolumn{10}{r}{\emph{Continued on next page}} \\
\bottomrule
\endfoot
\bottomrule
\endlastfoot
  3863.29 &   3862.60 &   11.999 $\pm$   2.181&  10.700 $^{  +2.300}_{  -2.300}$ & Si II      &  V1    &  3p2 2D  &   4p 2P*     &    4   &     2 \\
  3869.44 &   3868.75 &   25.938 $\pm$   2.070&  27.100 $^{  +2.300}_{  -2.500}$ & [Ne III]   &  F1    &  2p4 3P  &   2p4 1D     &    5   &     5 \\
  3882.88 &   3882.19 &    8.490 $\pm$   1.718&   6.320 $^{  +1.820}_{  -1.870}$ & O II       &  V12   &  3p 4D*  &   3d 4D      &    8   &     8 \\
  3889.34 &   3888.65 &   12.876 $\pm$   1.179&  14.200 $^{  +1.300}_{  -1.400}$ & He I       &  V2    &  2s 3S   &   3p 3P*     &    3   &     9 \\
  3908.16 &   3907.46 &    2.662 $\pm$   0.759&   3.152 $^{  +0.822}_{  -0.819}$ & O II       &  V11   &  3p 4D*  &   3d 4P      &    6   &     6 \\
  3968.17 &   3967.46 &    9.202 $\pm$   1.120&   8.170 $^{  +1.210}_{  -1.210}$ & [Ne III]   &  F1    &  2p4 3P  &   2p4 1D     &    3   &     5 \\
  3970.78 &   3970.07 &   13.609 $\pm$   1.120&  13.400 $^{  +1.200}_{  -1.300}$ & H I        &  H7    &  2p+ 2P* &   7d+ 2D     &    8   &    98 \\
  4026.80 &   4026.08 &    3.576 $\pm$   0.553&   3.553 $^{  +0.593}_{  -0.603}$ & N II       &  V39b  &  3d 3F*  &   4f 2[5]    &    7   &     9 \\
  4042.03 &   4041.31 &    1.154 $\pm$   0.518&   1.659 $^{  +0.545}_{  -0.546}$ & N II       &  V39b  &  3d 3F*  &   4f 2[5]    &    9   &    11 \\
  4068.67 &   4067.94 &    1.738 $\pm$   0.666&   2.129 $^{  +0.702}_{  -0.737}$ & C III      &  V16   &  4f 3F*  &   5g 3G      &    5   &     7 \\
  4084.63 &   4083.90 &    4.964 $\pm$   0.772&   4.373 $^{  +0.809}_{  -0.827}$ & O II       &  V48b  &  3d 4F   &   4f G4*     &    6   &     8 \\
  4102.47 &   4101.74 &   21.392 $\pm$   1.483&  24.600 $^{  +1.500}_{  -1.600}$ & H I        &  H6    &  2p+ 2P* &   6d+ 2D     &    8   &    72 \\
  4157.27 &   4156.53 &    2.172 $\pm$   0.573&   2.208 $^{  +0.598}_{  -0.596}$ & O II       &  V19   &  3p 4P*  &   3d 4P      &    6   &     4 \\
  4169.71 &   4168.97 &    8.103 $\pm$   0.766&   8.698 $^{  +0.793}_{  -0.873}$ & He I       &  V52   &  2p 1P*  &   6s 1S      &    3   &     1 \\
  4200.58 &   4199.83 &    2.369 $\pm$   0.514&   2.564 $^{  +0.542}_{  -0.546}$ & He II      &  4.11  &  4f+ 2F* &   11g+ 2G    &   32   &   242 \\
  4220.12 &   4219.37 &    3.141 $\pm$   0.614&   2.959 $^{  +0.653}_{  -0.646}$ & Ne II      &  V52a  &  3d 4D   &   4f 2[4]*   &    8   &     8 \\
  4254.62 &   4253.86 &    1.975 $\pm$   0.513&   2.377 $^{  +0.541}_{  -0.536}$ & O II       &  V101  &  3d 2G   &   4f 2[5]*   &   10   &    10 \\
  4267.91 &   4267.15 &    1.348 $\pm$   0.403&   1.757 $^{  +0.425}_{  -0.430}$ & C II       &  V6    &  3d 2D   &   4f 2F*     &   10   &    14 \\
  4282.08 &   4281.32 &    3.745 $\pm$   0.569&   2.804 $^{  +0.612}_{  -0.603}$ & O II       &  V53b  &  3d 4P   &   4f D2*     &    6   &     6 \\
  4341.24 &   4340.47 &   37.418 $\pm$   1.424&  40.200 $^{  +1.400}_{  -1.500}$ & H I        &  H5    &  2p+ 2P* &   5d+ 2D     &    8   &    50 \\
  4346.32 &   4345.55 &    7.697 $\pm$   1.444&  10.400 $^{  +1.500}_{  -1.500}$ & O II       &  V2    &  3s 4P   &   3p 4P*     &    4   &     2 \\
  4363.99 &   4363.21 &    5.238 $\pm$   0.574&   4.825 $^{  +0.572}_{  -0.648}$ & [O III]    &  F2    &  2p2 1D  &   2p2 1S     &    5   &     1 \\
  4367.67 &   4366.89 &    4.670 $\pm$   0.502&   4.987 $^{  +0.501}_{  -0.557}$ & N III      &  V2    &  3s 4P   &   3p 4P*     &    6   &     4 \\
  4413.90 &   4413.11 &    2.143 $\pm$   0.463&   2.172 $^{  +0.503}_{  -0.467}$ & Ne II      &  V57c  &  3d 4F   &   4f 1[3]*   &    4   &     6 \\
  4429.31 &   4428.52 &    3.250 $\pm$   0.784&   3.508 $^{  +0.824}_{  -0.817}$ & Ne II      &  V61b  &  3d 2D   &   4f 2[3]*   &    6   &     8 \\
  4466.45 &   4466.42 &    2.245 $\pm$   0.597&   2.430 $^{  +0.601}_{  -0.624}$ & O II       &  V86b  &  3d 2P   &   4f D2*     &    4   &     6 \\
  4514.89 &   4514.86 &    1.125 $\pm$   0.235&   1.049 $^{  +0.245}_{  -0.235}$ & N III      &  V3    &  3s' 4P* &   3p' 4D     &    6   &     8 \\
  4541.62 &   4541.59 &    4.152 $\pm$   0.503&   4.376 $^{  +0.500}_{  -0.564}$ & He II      &  4.9   &  4f+ 2F* &   9g+ 2G     &   32   &   162 \\
  4552.56 &   4552.53 &    1.708 $\pm$   0.478&   2.103 $^{  +0.493}_{  -0.489}$ & N II       &  V58a  &  3d 1F*  &   4f 2[4]    &    7   &     9 \\
  4685.71 &   4685.68 &  109.631 $\pm$   8.349& 110.000 $^{  +8.000}_{  -8.000}$ & He II      &  3.4   &  3d+ 2D  &   4f+ 2F*    &   18   &    32 \\
  4711.40 &   4711.37 &   12.518 $\pm$   1.284&  12.500 $^{  +1.300}_{  -1.300}$ & [Ar IV]    &  F1    &  3p3 4S* &   3p3 2D*    &    4   &     6 \\
  4724.18 &   4724.89 &    0.501 $\pm$   0.463&   0.000 $^{  +0.000}_{  -0.452}$ & [Ne IV]    &  F1    &  2p3 2D* &   2p3 2P*    &    4   &     4 \\
  4740.20 &   4740.17 &    7.183 $\pm$   0.720&   8.278 $^{  +0.701}_{  -0.766}$ & [Ar IV]    &  F1    &  3p3 4S* &   3p3 2D*    &    4   &     4 \\
  4769.43 &   4769.40 &    1.838 $\pm$   0.342&   1.196 $^{  +0.349}_{  -0.339}$ & [Fe III]   &  F 3   &          &              &    0   &     0 \\
  4861.36 &   4861.33 &   98.409 $\pm$   5.838& 100.000 $^{  +5.900}_{  -5.900}$ & H I        &  H4    &  2p+ 2P* &   4d+ 2D     &    8   &    32 \\
  4958.94 &   4958.91 &   92.697 $\pm$   5.533&  92.500 $^{  +5.500}_{  -5.500}$ & [O III]    &  F1    &  2p2 3P  &   2p2 1D     &    3   &     5 \\
  5006.87 &   5006.84 &  284.817 $\pm$  13.799& 280.000 $^{  +14.00}_{  -14.00}$ & [O III]    &  F1    &  2p2 3P  &   2p2 1D     &    5   &     5 \\
  5046.71 &   5047.74 &    1.488 $\pm$   0.153&   1.325 $^{  +0.142}_{  -0.160}$ & He I       &  V47   &  2p 1P*  &   4s 1S      &    3   &     1 \\
  5269.32 &   5270.40 &    0.552 $\pm$   0.179&   0.872 $^{  +0.174}_{  -0.170}$ & [Fe III]   &  F1    &  3d6 5D  &   3d6 3P2    &    7   &     5 \\
  5341.29 &   5342.38 &    0.566 $\pm$   0.059&   0.539 $^{  +0.055}_{  -0.061}$ & C II       &        &  4f 2F*  &   7g 2G      &   14   &    18 \\
  5410.41 &   5411.52 &    7.603 $\pm$   0.516&   7.694 $^{  +0.495}_{  -0.529}$ & He II      &  4.7   &  4f+ 2F* &   7g+ 2G     &   32   &    98 \\
  5516.53 &   5517.66 &    0.645 $\pm$   0.098&   0.657 $^{  +0.088}_{  -0.102}$ & [Cl III]   &  F1    &  2p3 4S* &   2p3 2D*    &    4   &     6 \\
  5536.47 &   5537.60 &    0.479 $\pm$   0.109&   0.518 $^{  +0.104}_{  -0.105}$ & [Cl III]   &  F1    &  2p3 4S* &   2p3 2D*    &    4   &     4 \\
  5576.20 &   5577.34 &    4.415 $\pm$   0.474&   3.909 $^{  +0.436}_{  -0.491}$ & [O I]      &  F3    &  2p4 1D  &   2p4 1S     &    5   &     1 \\
  5665.92 &   5666.63 &    0.771 $\pm$   0.164&   0.667 $^{  +0.155}_{  -0.154}$ & N II       &  V3    &  3s 3P*  &   3p 3D      &    3   &     5 \\
  5695.21 &   5695.92 &    0.715 $\pm$   0.143&   0.625 $^{  +0.133}_{  -0.133}$ & C III      &  V2    &  3p 1P*  &   3d 1D      &    3   &     5 \\
  5753.88 &   5754.60 &    0.193 $\pm$   0.042&   0.198 $^{  +0.038}_{  -0.041}$ & [N II]     &  F3    &  2p2 1D  &   2p2 1S     &    5   &     1 \\
  5800.79 &   5801.51 &    1.589 $\pm$   0.216&   1.123 $^{  +0.204}_{  -0.205}$ & C IV       &  V1    &  3s 2S   &   3p 2P*     &    2   &     4 \\
  5867.27 &   5868.00 &    0.688 $\pm$   0.152&   0.726 $^{  +0.139}_{  -0.146}$ & [Kr IV]    &        &  4p3 4S  &   3d3 2G     &    4   &     4 \\
  5874.93 &   5875.66 &    0.902 $\pm$   0.143&   0.772 $^{  +0.137}_{  -0.134}$ & He I       &  V11   &  2p 3P*  &   3d 3D      &    9   &    15 \\
  5978.23 &   5978.97 &    0.591 $\pm$   0.086&   0.401 $^{  +0.080}_{  -0.080}$ & S III      &  V4    &          &              &    0   &     0 \\
  6035.95 &   6036.70 &    0.731 $\pm$   0.093&   0.418 $^{  +0.088}_{  -0.087}$ & He II      &  5.21  &  5g+ 2G  &   21h+ 2H*   &   50   &   882 \\
  6073.34 &   6074.10 &    0.464 $\pm$   0.099&   0.351 $^{  +0.094}_{  -0.092}$ & He II      &  5.20  &  5g+ 2G  &   20h+ 2H*   &   50   &   800 \\
  6101.07 &   6101.83 &    0.351 $\pm$   0.084&   0.454 $^{  +0.078}_{  -0.078}$ & [K IV]     &  F1    &  3p4 3P  &   3d4 1D     &    5   &     5 \\
  6117.44 &   6118.20 &    0.239 $\pm$   0.075&   0.369 $^{  +0.070}_{  -0.069}$ & He II      &  5.19  &  5g+ 2G  &   19h+ 2H*   &   50   &   722 \\
  6169.92 &   6170.69 &    0.536 $\pm$   0.086&   0.442 $^{  +0.079}_{  -0.079}$ & He II      &  5.18  &  5g+ 2G  &   18h+ 2H*   &   50   &   648 \\
  6233.89 &   6233.80 &    0.475 $\pm$   0.056&   0.474 $^{  +0.053}_{  -0.054}$ & He II      &  5.17  &  5g+ 2G  &   17h+ 2H*   &   50   &   578 \\
  6406.39 &   6406.30 &    0.612 $\pm$   0.058&   0.503 $^{  +0.053}_{  -0.059}$ & He II      &  5.15  &  5g+ 2G  &   15h+ 2H*   &   50   &   450 \\
  6527.21 &   6527.11 &    2.066 $\pm$   0.248&   1.781 $^{  +0.221}_{  -0.253}$ & He II      &  5.14  &  5g+ 2G  &   14h+ 2H*   &   50   &   392 \\
  6562.87 &   6562.77 &  283.836 $\pm$  17.889& 254.000 $^{ +12.000}_{  -9.000}$ & H I        &  H3    &  2p+ 2P* &   3d+ 2D     &    8   &    18 \\
  6683.30 &   6683.20 &    1.006 $\pm$   0.094&   0.856 $^{  +0.086}_{  -0.095}$ & He II      &  5.13  &  5g+ 2G  &   13h+ 2H*   &   50   &   338 \\
  6716.54 &   6716.44 &    0.543 $\pm$   0.139&   0.511 $^{  +0.127}_{  -0.130}$ & [S II]     &  F2    &  2p3 4S* &   2p3 2D*    &    4   &     6 \\
  6730.92 &   6730.82 &    0.837 $\pm$   0.153&   0.610 $^{  +0.139}_{  -0.135}$ & [S II]     &  F2    &  2p3 4S* &   2p3 2D*    &    4   &     4 \\
  7006.92 &   7005.67 &    4.287 $\pm$   0.353&   3.378 $^{  +0.331}_{  -0.366}$ & [Ar V]     &  F1    &  3s2 3P  &   3s2 1D     &    5   &     5 \\
  7066.52 &   7065.25 &    0.271 $\pm$   0.076&   0.249 $^{  +0.067}_{  -0.069}$ & He I       &  V10   &  2p 3P*  &   3s 3S      &    9   &     3 \\
  7137.08 &   7135.80 &    2.759 $\pm$   0.229&   2.146 $^{  +0.212}_{  -0.236}$ & [Ar III]   &  F1    &  3p4 3P  &   3p4 1D     &    5   &     5 \\
  7178.79 &   7177.50 &    1.166 $\pm$   0.094&   0.902 $^{  +0.089}_{  -0.098}$ & He II      &  5.11  &  5g+ 2G  &   11h+ 2H*   &   50   &   242 \\
  7237.49 &   7236.19 &    0.186 $\pm$   0.156&   0.409 $^{  +0.133}_{  -0.138}$ & C II       &  V3    &  3p 2P*  &   3d 2D      &    4   &     6 \\
  7264.06 &   7262.76 &    0.371 $\pm$   0.100&   0.265 $^{  +0.086}_{  -0.090}$ & [Ar IV]    &  F2    &  3p3 2D* &   3p3 2P*    &    4   &     2 \\
  7320.23 &   7319.45 &    0.046 $\pm$   0.100&   0.000 $^{  +0.000}_{  -0.085}$ & [O II]     &  F2    &  2p3 2D* &   2p3 2P*    &    6   &     2 \\
  
    \label{tab:A2}
\end{longtable}

\twocolumn{

\balance


\bibliographystyle{apj} 
\bibliography{references_new}
}

\end{document}